\documentclass[acmsmall,screen,nonacm]{acmart}

\usepackage{amsmath,amsfonts}
\usepackage{algorithmic}
\usepackage{graphicx}
\usepackage[dvipsnames]{xcolor}
\usepackage[italic]{mathastext}
\usepackage{libertine}
\usepackage[T1]{fontenc}
\usepackage{textcomp}
\usepackage[all]{nowidow}
\usepackage[keeplastbox]{flushend}
\usepackage{fancyhdr}
\usepackage{float}

\usepackage{color} %
\usepackage{hyperref} %

\usepackage{graphicx} %
\usepackage{listings}
\usepackage{booktabs}
\usepackage{tabularx}
\usepackage{array}
\usepackage{outlines}
\usepackage{boldline}
\usepackage{graphicx}
\graphicspath{{figs/}}
\usepackage{listings}
\usepackage{makecell}
\usepackage{mathtools}
\usepackage{multirow}
\usepackage{xspace}
\usepackage{pifont}
\usepackage{booktabs}
\usepackage{pifont}

\usepackage{xurl}

\usepackage{breakurl}
\usepackage{extarrows}
\usepackage{textcomp}
\usepackage{xparse}   %
\usepackage{bigdelim}
\usepackage[linesnumbered,ruled,vlined]{algorithm2e}
\usepackage{appendix}
\usepackage{color}
\usepackage{tikz}
\usepackage{outlines}
\usepackage{threeparttable}
\usepackage{subcaption}
\usepackage{wrapfig}
\usepackage{textcomp}
\usepackage{subcaption}
\usepackage[table,xcdraw,svgnames,dvipsnames,HTML]{xcolor}
\usepackage[all]{nowidow}
\usepackage[keeplastbox]{flushend}
\usepackage{fancyhdr}

\newcolumntype{C}[1]{>{\centering\arraybackslash}m{#1}}
\newcolumntype{R}[1]{>{\raggedleft\arraybackslash}p{#1}}
\newcolumntype{L}[1]{>{\raggedright\arraybackslash}p{#1}}
\newcommand{\raggedcentercell}[1]{%
  \begin{minipage}[c][\totalheight][c]{\linewidth}%
  \raggedright\arraybackslash #1%
  \end{minipage}%
}

\hypersetup{colorlinks=false, pdfborder={0 0 0}}

\begin{document}

\newcommand{\citeXXX}[1]{\textcolor{red}{[X]}\xspace}
\newcommand{\XXX}{\textcolor{red}{[XXX]}\xspace}
\newcommand{\YYY}{\textcolor{red}{[YYY]}\xspace}

\newcommand{\sysname}{FerroScope\xspace}

\newcommand{\NUMMODELS}{four\xspace}
\newcommand{\NUMAPPS}{five\xspace}

\newcommand{\AVGCOMPILESUCCESSGPTFOUR}{{89.8\%}\xspace}
\newcommand{\AVGCOMPILESUCCESSGPTTHREEFIVE}{{70.5\%}\xspace}
\newcommand{\AVGCOMPILESUCCESSGPTMINI}{{83.0\%}\xspace}
\newcommand{\AVGCOMPILESUCCESSCLAUDE}{{48.3\%}\xspace}

\newcommand{\AVGUNSAFEGPTFOUR}{{12.5\%}\xspace}
\newcommand{\AVGUNSAFEGPTTHREEFIVE}{{32.0\%}\xspace}
\newcommand{\AVGUNSAFEGPTMINI}{{19.7\%}\xspace}
\newcommand{\AVGUNSAFECLAUDE}{{12.9\%}\xspace}

\newcommand{\AVGSCOREGPTFOUR}{{69.9\%}\xspace}
\newcommand{\AVGSCOREGPTTHREEFIVE}{{70.2\%}\xspace}
\newcommand{\AVGSCOREGPTMINI}{{72.8\%}\xspace}
\newcommand{\AVGSCORECLAUDE}{{77.8\%}\xspace}

\newcommand{\AVGCOVERAGEFOUR}{{58.3\%}\xspace}
\newcommand{\AVGCOVERAGETHREEFIVE}{{67.2\%}\xspace}
\newcommand{\AVGCOVERAGEMINI}{{60.9\%}\xspace}
\newcommand{\AVGSCOVERAGECLAUDE}{{67.1\%}\xspace}

\newcommand{\AVGCOVERAGELIBCSV}{{77.4\%}\xspace}
\newcommand{\AVGCOVERAGEOPTIPNG}{{51.6\%}\xspace}
\newcommand{\AVGCOVERAGEURLPARSER}{{62.9\%}\xspace}
\newcommand{\AVGSCOVERAGELIBBMP}{{68.5\%}\xspace}
\newcommand{\AVGSCOVERAGEUC}{{87.9\%}\xspace}

\newcommand{\SYMBOLICEXECTIMEOUT}{{180 minutes}\xspace}
\newcommand{\AVGCOVERAGE}{{62.9\%}\xspace}
\newcommand{\rustminus}{{rust$--$}\xspace}
\newcommand{\rmm}{{\mbox{Rust-{}-}\xspace}}

\newcommand{\purge}[1]{{\textbf{#1}}}
\renewcommand{\purge}[1]{}
\newcommand{\APPS}{9\xspace}
\newcommand{\AVGOVERHEAD}{5.45}
\newcommand{\AVGPTSIMP}{13.15}
\newcommand{\MAXPTSIMP}{1.25}
\newcommand{\PYLOC}{{139}\xspace}
\newcommand{\SYMEXECTIMEOUT}{\textcolor{blue}{\textbf{10800}}\xspace}

\renewcommand{\citeXXX}{}
\renewcommand{\XXX}{}
\renewcommand{\YYY}{}

\newcommand{\intel}{Intel\textsuperscript{\textregistered}\xspace}
\newcommand{\intels}{Intel's\textsuperscript{\textregistered}\xspace}

\newcommand{\critarg}{Precision Critical Arguments\xspace}
\newcommand{\LIH}{optimistic analysis\xspace}
\newcommand{\LINH}{baseline analysis\xspace}
\newcommand{\LIHMV}{optimistic MV\xspace}
\newcommand{\LINHMV}{fallback MV\xspace}
\newcommand{\Addrof}{\emph{Addr-Of}\xspace}
\newcommand{\Deref}{\emph{Load}\xspace}
\newcommand{\Store}{\emph{Store}\xspace}
\newcommand{\Copy}{\emph{Copy}\xspace}
\newcommand{\FieldOf}{\emph{Field-Of}\xspace}

\def\Snospace~{\S{}}
\def\sectionautorefname{\Snospace}
\def\subsectionautorefname{\Snospace}
\def\subsubsectionautorefname{\Snospace}

\definecolor{lightergray}{HTML}{dedede}

\newcommand\para[1]{\vspace{0.2em} \noindent \textbf{#1}}
\newcommand\parasmall[1]{\vspace{1em} \noindent \textbf{#1}}

\newcommand{\squishenumerate}{
   \begin{enumerate}
       { \setlength{\itemsep}{0pt}  \setlength{\parsep}{0pt}
      \setlength{\topsep}{1pt}       
      \setlength{\partopsep}{0pt}
      \setlength{\leftmargin}{1em} 
      \setlength{\labelwidth}{0em}
      \setlength{\labelsep}{0.0em} } }

\newcommand{\squishenumerateend}{
    \end{enumerate}  }

\newcommand{\squishlist}{
   \begin{list}{$\bullet$}
    { \setlength{\itemsep}{0pt}      \setlength{\parsep}{3pt}
      \setlength{\topsep}{3pt}       \setlength{\partopsep}{0pt}
      \setlength{\leftmargin}{1em} \setlength{\labelwidth}{1em}
      \setlength{\labelsep}{0.5em} } }

\newcommand{\squishlisttwo}{
   \begin{list}{$\bullet$}
    { \setlength{\itemsep}{0pt}    \setlength{\parsep}{0pt}
      \setlength{\topsep}{0pt}     \setlength{\partopsep}{0pt}
      \setlength{\leftmargin}{2em} \setlength{\labelwidth}{1.5em}
      \setlength{\labelsep}{0.5em} } }

\newcommand{\squishend}{
    \end{list}  }

\title{Cross-Layer Characterization and AI-Assisted Diagnosis of C-to-Rust Performance}

\author{Dalton Davis}
\affiliation{%
  \institution{University of California, Davis}
  \city{Davis}\state{CA}\country{USA}}
\email{drdavis@ucdavis.edu}

\author{Kunal Pai}
\affiliation{%
  \institution{University of California, Los Angeles}
  \city{Los Angeles}\state{CA}\country{USA}}
\email{kunpai@ucla.edu}

\author{Tapti Palit}
\affiliation{%
  \institution{University of California, Davis}
  \city{Davis}\state{CA}\country{USA}}
\email{tpalit@ucdavis.edu}

\begin{abstract}

Migrating C code to Rust can significantly improve software
security, but understanding the performance implications of
such migrations remains challenging. While Rust is often
claimed to match C's performance, runtime safety checks,
compiler optimizations, and hardware microarchitecture
interact in ways that complicate performance diagnosis.
We first conduct a systematic cross-layer characterization
of 12 benchmarks comprising C implementations and their
Rust counterparts. Our study reveals that runtime safety
checks do not necessarily dominate performance overhead,
as compiler and microarchitectural optimizations often
mitigate their costs. Across 8 of the 12 benchmarks,
at least one Rust implementation performs within 10\%
of its C counterpart or outperforms it.
Motivated by these findings, we introduce \sysname,
an AI-assisted performance debugger that integrates
source-level analysis, compiler optimization remarks,
dynamic instruction traces, and hardware performance
counters to diagnose performance differences between
C and Rust. 
\sysname reproduces our manual findings
while providing more precise explanations in several cases,
including identifying compiler optimizations inhibited
by redundant bounds checks. Guided by these diagnoses,
we develop an LLVM compiler pass that hoists redundant
bounds checks through guarded loop versioning while
preserving Rust's panic behavior. The pass enables
downstream compiler optimizations, including loop
unrolling and conditional-move conversion, reducing
Selection Sort's overhead from 2.95$\times$ to
1.00$\times$ and improving Merge Sort from
1.43$\times$ to 0.91$\times$ relative to C.
Together, our results demonstrate how cross-layer
performance characterization and AI-assisted diagnosis
can guide safety-preserving compiler optimizations
for efficient C-to-Rust migration.

\end{abstract}

\maketitle

\section{Introduction}

Rust~\cite{rust} is a memory-safe, systems programming language that is increasingly gaining
popularity.  
Unlike 
other memory-safe programming languages such as 
Java~\cite{java-gc} and Go~\cite{go-gc}, Rust does not rely on garbage collection~\cite{rust-ownership} 
to ensure temporal safety. Instead, the Rust compiler
ensures that the Rust code is free from temporal safety
vulnerabilities
at compile-time. Performing these
checks at compile time allows Rust to 
mitigate the performance overhead typically associated
with temporal memory safety. 
Due to Rust's memory safety
and performance characteristics, system software such as 
operating systems~\cite{tockos, narayanan2020redleaf,redoxos,boos2020theseus} and 
device drivers~\cite{e1000-crate, corbet-rust-linux} 
are being increasingly 
written in Rust. 
Recently, the Linux kernel has officially
added support for Rust code~\cite{ars-technica-linux-rust,kernel-rust-docs,linux-magazine-initial-rust}. 
Similarly, parts of web browsers~\cite{firefox-rust}
and web servers~\cite{nginx-rust} 
are being
rewritten in Rust. 

However, not all of 
Rust's safety checks happen at compile time. To guarantee
spatial safety, Rust abstractions, such as \emph{vectors}
and \emph{strings}, perform runtime checks to detect
out-of-bounds memory accesses. Rust also
supports run-time-checked pointers, including
reference-counted pointers (\texttt{RC}),
and types that support interior mutability safely 
(\texttt{RefCells}). 
The performance overhead for the runtime safety checks in 
Safe Rust code stems from many sources---some obvious, and some
not so obvious. The obvious source of performance overhead
is the cost of executing the additional \emph{machine
instructions} for the check itself. 
But there exist other non-obvious 
overhead sources, too. For
example, inserting the runtime checks can
potentially disrupt compiler optimizations. 
Bounds checks are essentially assembly-level \emph{branch instructions}
and inserting these branches can change how the compiler optimizes the surrounding
code. Furthermore, not all of Rust's language features
impeded compiler optimization. Rust's ownership model can also \emph{improve}
compiler optimizations in ways that partially offset these costs. The borrow checker's
guarantee that no mutable reference coexists with any other reference to the same
memory allows the compiler to enable more aggressive alias analysis and
optimization~\cite{villani2025treeborrows}. 

Rust's safety checks also impact the hardware 
microarchitectural
characteristics. The increase in the number of branch instructions
resulting from the memory safety checks
can
put pressure on the hardware branch predictor. This could potentially
increase the percentage of \emph{branch misses} and stall the CPU
pipeline, increasing the runtime overhead. 
However, on the other hand, if Rust's memory-safety checks result
in \emph{predictable} branch outcomes
and branch targets, then the hardware branch predictor
can mitigate much of the associated overhead. Similarly,
if the memory safety checks require loading the object's
bounds information from the DRAM,
 the performance overhead
would be masked if hardware caches can capture
this bounds information, thus reducing
the data access latency.

These factors complicate the estimation
of the performance of C code migrated to Rust. 
This presents a challenge for 
software developers who are considering rewriting
their legacy C codebases to Rust. 
While it is clear that Rust would outperform other garbage-collected
languages, there is less consensus surrounding its performance in comparison
to C code.
Performance results for Rust ports remain nuanced and vary by project. For 
instance, while the Rust implementation of \texttt{zlib}~\cite{libz-rs} 
reportedly \emph{outperforms} its C baseline, the 
Rust port of GNU 
CoreUtils~\cite{rust-coreutils} initially \emph{underperformed}. Only after subsequent targeted optimizations, core utilities such as \texttt{wc},
and 
\texttt{ls}
match or 
exceed the performance of their original C counterparts~\cite{larabel2025rustcoreutils}.
This lack of consensus also prevails in research works measuring C and Rust
performance. Some previous works~\cite{cui2024finding, zhang2022towards} have found
that the memory-safety checks contribute to significant overheads, reaching up
to 2.49$\times$ overhead. On the other hand, other works~\cite{popescu2021safer} 
have found that Rust code can outperform C code for some libraries. More recent 
comparisons~\cite{mihajlovic2026rustcpp, wells2026rustcpp} show that 
well-optimized Rust implementations typically stay within 5--10\% of C/C++ 
performance on average, with wins in some workloads and small losses in others.
However, the reasons
for this behavior have not yet been analyzed systematically,
and we aim to fill this gap. 
Confounding matters further, different rewriting strategies
can have different performance characteristics---from
line-by-line C-to-Rust rewrites to a
complete redesign and 
re-implementation of the program functionalities in Rust.

We introduce a framework for characterizing
and diagnosing performance differences between 
C and migrated Rust code. Using this framework
we demonstrate how
the resulting workload characterization
can guide compiler
optimizations. Our work proceeds from
a systematic characterization of C-to-Rust performance
to an AI-assisted performance debugger, and finally to
a compiler optimization informed by the identified
performance bottlenecks.

\para{Cross-Layer Performance Characterization Study.}
We first perform a systematic characterization
of manually translated Rust code
that decomposes language-level safety costs into
compiler and microarchitectural factors.
Our study includes 12 micro- and macro-benchmarks, comprising
original C implementations and their manual Rust 
translations.
Where applicable, we evaluate alternate Rust implementations
to investigate the impact of implementation strategies on
runtime performance.
The microbenchmarks include searching and sorting algorithms
that operate on dynamically sized buffers and incur frequent
runtime bounds checks.
The macrobenchmarks include \emph{Skip List}~\cite{vadrevu2024cannot}
and \emph{PageRank}~\cite{page1999pagerank} implementations,
and the \emph{LibCsv}~\cite{noauthor_libcsv_nodate}
and \emph{u8c}~\cite{u8c_lib} libraries.

Our characterization reveals that, contrary to common belief,
Rust's safety mechanisms do not \emph{always} dominate runtime
overheads.
Instead, compiler and microarchitectural optimizations often
hide these costs.
Across 8 of the 12 evaluated benchmarks, at least one Rust
implementation achieved performance within a 10\% margin
of the corresponding C baseline or exceeded it.
Our analysis reveals a three-tier interaction between
language-level safety mechanisms, compiler optimizations,
and hardware microarchitecture, demonstrating that performance
differences cannot be attributed solely to runtime safety checks.
We further observe that differences in compiler optimizations
between C and Rust can introduce unexpected slowdowns.
These findings demonstrate that understanding performance
regressions in C-to-Rust migration requires reasoning across
multiple abstraction layers, from language-level constructs
to compiler-generated instructions and their interactions
with hardware microarchitecture.

\para{AI-assisted Performance Debugger.} Motivated by the complexity of these cross-layer
interactions, we develop \sysname---an 
AI-Assisted cross-layer performance
debugger.  \sysname uses a multi-stage
diagnostic pipeline that combines source-level analysis,
LLVM intermediate representations~\cite{llvmlangref}, compiler optimization
remarks~\cite{llvm-remarks}, dynamic instruction traces, and hardware
performance counters~\cite{perf}. By correlating evidence across
these layers, \sysname identifies potential causes of
performance differences and produces evidence-supported
explanations. We evaluate \sysname against our manual
characterization and find that it reproduces all of
the manually established explanations while sharpening
the diagnosis in many cases. For example,
\sysname leverages the compiler optimization
remarks and identifies compiler optimizations inhibited
by redundant bounds checks, as well as, distinguishes
implementation-level overhead from micro-architectural
effects.

\para{Diagnosis-Guided Compiler Optimization.}
We further demonstrate how 
the performance debugging results
can be used to inform compiler optimizations. We
focus on the \texttt{rustc} compiler's treatment
of the runtime spatial safety checks. We observed
that performing these safety checks inside hot loops
was expensive and also that, while in many cases it
was theoretically possible to \emph{hoist} these
checks to the loop preheader, 
the current compiler passes are unable to do so. 
We develop an LLVM~\cite{llvm} 
compiler pass that hoists provably
redundant bounds checks out of loops through guarded
loop versioning. The pass preserves Rust's 
observable panic behavior by retaining the 
original loop as a
fallback when the preheader checks cannot establish
safety. Removing redundant checks enables subsequent
LLVM optimizations that were previously inhibited.
Our evaluation shows that the pass reduces Selection
Sort's overhead from 2.95$\times$ to 1.00$\times$,
and improves Merge Sort from 1.43$\times$
to 0.91$\times$ relative to C.

\section{Background}

\subsection{Rust Memory Safety}

\para{Compile-time Checks.} 
Rust 
achieves memory safety through a combination of compile-time and runtime checks.
Rust enforces a strict ownership system at compile time, encompassing three core
concepts, namely ownership, lifetime, and borrowing.
In Rust, every object has a single 
owner, and when that owner goes out of scope, the object is automatically dropped and its memory freed. 
References can borrow an object's value, but the compiler ensures these references never outlive the 
owner. 
Additionally, the Rust Borrow Checker also enforces the \emph{Mutability-xor-aliasing} rule,
which ensures that heap variables can have either multiple immutable references or a single
mutable reference, but never 
both simultaneously. Together, these rules provide strong temporal safety guarantees
against
\emph{Use-After-Free} and \emph{Double-Free} bugs, and prevent data races.

\para{Runtime Checks.} 
Rust enforces spatial safety through runtime bounds checking. 
Every time a vector or slice element is accessed by index, Rust inserts a check to verify that 
this index is within the object range.
These checks eliminate spatial safety bugs but can incur some performance overhead. 
Consider the simplified memory safety check
for a slice indexing operation shown in \autoref{fig:safetycheck}. The index operation is
performed by the \texttt{core::slice::index::SliceIndex<[T]>::index} function, in the
Rust standard library. As the 
assembly instructions show, this check involves (i) loading the slice's \emph{size} into the 
\texttt{rdi} register, (ii) performing a \texttt{test} operation, and finally (iii) performing
a conditional branch operation using the \texttt{jne} instruction. As we will
discuss next, these instructions
interact with compiler and microarchitectural optimizations, thus
impacting runtime performance. 

\texttt{RefCell<T>}
performs borrow checking at runtime by tracking active borrows, instead of enforcing
Rust's borrowing rules at compile time. 
This allows the programmer to explicitly call the \texttt{borrow\_mut()} or \texttt{borrow()} functions to borrow the interior value. 
Attempting to obtain a mutable borrow while any other borrow is active causes a panic. 
Rust also provides reference-counted pointers through \texttt{Rc<T>}, which enable shared ownership when compile-time ownership rules are too restrictive. 
The strong reference count of an \texttt{Rc} is incremented and decremented by \texttt{clone()} and \texttt{drop()}, 
respectively.
The underlying value is dropped when the strong reference count reaches zero, 
while the allocation is retained until the weak reference count also reaches zero.
\texttt{Rc<T>} and \texttt{RefCell<T>} are commonly used to implement shared mutable data structures, such as graphs and linked lists.

\begin{figure}
    \centering
    \includegraphics[width=0.98\columnwidth]{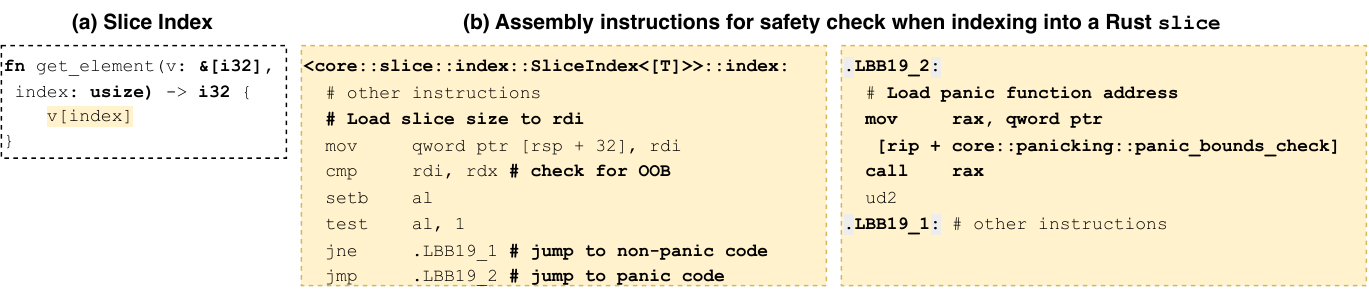}
    \caption{Rust Code and Assembly Instructions for Simplified Rust Runtime Safety Check.}
    \label{fig:safetycheck}
    \vspace{-1em}
\end{figure}

\subsection{Compiler Toolchains and Optimizations}
\label{sec:bg-compiler}

The Rust compiler (\texttt{rustc}) uses a multi-stage compilation pipeline, which
lowers the Rust code to assembly code.
Both \texttt{rustc} and \texttt{clang} use the same LLVM backend for low-level optimizations.
Thus, a significant part of the code generation is similar across Rust and C.

Multiple compiler optimizations can impact the runtime performance of Rust's safety checks.
For example, for the slice index check code provided in \autoref{fig:safetycheck}, one of the simplest compiler
optimizations is to inline the \texttt{index} function. 
In cases where the code contains multiple index operations into 
the same slice,
the compiler might optimize loading the slice size from memory by promoting
the bounds value to a register and
retaining it between subsequent accesses, thus amortizing the slow memory access. And finally, the compiler
can choose to rewrite the two-instruction sequence of the conditional branch (\texttt{jne})
followed by the memory load (\texttt{mov rax, [panic\_bounds\_check]}) 
into a single conditional memory load (\texttt{cmov}) instruction, thus reducing the instruction count.
We discuss potential impacts of this optimization in \autoref{sec:compilereffects}.

\subsection{Microarchitectural Optimizations}
\label{sec:bg-micro}

Compilers optimize binaries to exploit microarchitectural structures such as caches, branch predictors, TLBs, and
hardware prefetchers.
However, language-level safety guarantees can disrupt the application of such optimizations.
Specifically, the additional memory loads introduced by safety checks (e.g., bounds checks in \autoref{fig:safetycheck}) can be absorbed if they hit in the L1 or L2 data caches, but can result in performance penalties if they evict frequently accessed data or increase instruction cache pressure.
Similarly, highly predictable conditional branches in bounds checks allow the hardware branch predictor to  absorb their cost.
Hardware prefetchers and TLBs further mitigate latency by anticipating future accesses and caching virtual-to-physical translations~\cite{gao2024scalable,ryoo2017rethinking}, respectively, though mispredictions or cache thrashing can negate these benefits.

\section{Characterization Methodology and Setup}

\subsection{Benchmarks}
\label{sec:design}

Our characterization consists of eight microbenchmarks and four application-
and library-level macrobenchmarks. Our benchmark selection is
motivated by three goals: (1) using workloads commonly found in
systems-level code or prior C-to-Rust transpilation studies~\cite{
emre_translating_2021,nitin2025c2saferrust,
zhang2023ownership,zhang2022towards},
(2) covering diverse memory access patterns (regular, semi-regular,
and irregular), and (3) stress-testing Rust's runtime safety checks.

\para{Microbenchmarks.}
Our microbenchmarks include sorting, searching, graph traversal,
and matrix multiplication algorithms with diverse access
patterns~\cite{docs-sched-cfs,linux-rbtree-doc,linux-vfs-dcache,
linux-generic-radix,gcc-summit-2006-treegion,
gcc-internals-control-dataflow,sqlite-btree-binary-search,
chetlur2014cudnn,uber:deepeta,beamer2015gap,shun2013ligra}.
Breadth-First Search (BFS) and Depth-First Search (DFS) operate
on adjacency matrices with 30,000 nodes, while Single-Source
Shortest Path (SSSP) uses an adjacency-list representation
with 30,000 nodes. Binary Search operates on a sorted array
of 20 million integers, while Merge Sort and Quicksort each
process arrays of 100 million integers. Selection Sort uses
200,000 integers, and Matrix Multiplication operates on
two-dimensional integer matrices of size $2000 \times 2000$.

\para{Application Benchmarks.}
To evaluate irregular, pointer-chasing access patterns, we include
Skip List~\cite{leveldb:github,redis,vadrevu2024cannot,
rocksdb:github} and PageRank~\cite{page1999pagerank,
beamer2015gap,shun2013ligra}. These represent linked and graph
structures commonly used in systems such as LevelDB and Redis.
The Skip List benchmark operates on a linked-node structure with
$N=5$ million nodes, performing $3N/2$ insert operations
(including $N/2$ updates), $3N/4$ removals, and $N$ searches,
with both successful and unsuccessful lookups.
We evaluate PageRank on the LiveJournal social network
graph~\cite{backstrom2006group,leskovec2009community},
which contains approximately 4 million nodes and 35 million
edges, represented using adjacency lists.

\para{Library Benchmarks.}
For control-intensive workloads with regular streaming access
patterns, we use two libraries. U8C~\cite{u8c_lib} is a UTF-8
library used in Rust crates such as \texttt{serde}~\cite{serde_repo}
and \texttt{ripgrep}~\cite{ripgrep_repo}. We benchmark its
\texttt{u8next} decoder, a DFA-based state machine with multiple
branches and state transitions, on a text file containing
approximately 2.8 million characters~\cite{UnicodeNormalizationTest}.
LibCSV~\cite{libcsv} is an ANSI-C CSV parsing library. We
benchmark its \texttt{csv\_parse} function, a single-pass,
byte-oriented state machine with branch-heavy control flow,
using the COVID-19 Open Research Dataset~\cite{
cord19kaggle,wang-etal-2020-cord}, containing approximately
1 million rows (1.65 GB).

\subsection{Evaluation Methodology}
As discussed in \autoref{sec:bg-compiler} and \autoref{sec:bg-micro}, the microarchitectural
characteristics of a workload depend on the
program-level data access patterns and also how the compiler optimizes the program.
To analyze the workload's 
microarchitectural characteristics,
\sysname uses Pin tools~\cite{pintool} to record the dynamic instruction composition
and the Hardware Performance Counters~\cite{perf}, discussed in \autoref{sec:bg-micro}, to record
the microarchitectural characteristics of the workload. 
In particular, \sysname records L1 and L2 data cache misses
to analyze (i) the cache locality of memory safety metadata
and (ii) changes in memory access optimizations caused by
Rust's abstractions and safety checks. It also records
hardware branch predictor misses to identify overhead
from unpredictable safety-check branches or increased
branch predictor pressure. Finally, \sysname uses a
Pin tool~\cite{pintool} to collect the dynamic instruction
mix, focusing on memory loads, stores, and conditional
branches that can potentially stall the CPU pipeline.

\subsection{Evaluation Setup}
\label{sec:eval}

\para{Hardware Details.}
We conducted our experiments on the CloudLab~\cite{Duplyakin+:ATC19} \texttt{c6525-25g} nodes 
in the Utah cluster for AMD and \texttt{sm110p} nodes in the Wisconsin cluster for Intel.
Both systems used Linux v5.15.0-168-generic.
The AMD node is a Dell PowerEdge C6525 server equipped with an AMD EPYC 7302P processor. 
The machine had 16 physical cores (32 threads with SMT) running at 3.0 GHz and a 155 W TDP. 
Each core has 32 KiB L1D and 32 KiB L1I caches, 512 KiB L2 cache, and 16 MiB L3 per CCX (128 MiB total). 
The Intel node is a Supermicro SuperServer 110P equipped with an Intel Xeon Silver 4314 processor
(Ice Lake-SP, Sunny Cove microarchitecture)~\cite{intel2021xeon4314}.
The machine had 16 physical cores (32 threads with Hyper-Threading) with a 2.4 GHz base clock
(3.4 GHz turbo) and a 135 W TDP.
Each core has 48 KiB L1D and 32 KiB L1I caches, 1.25 MiB L2 cache, 
and a non-inclusive shared 24 MiB L3 cache~\cite{intel2020icelakesp_hotchips}.

\para{Compilers.} All C benchmarks were compiled with the \texttt{clang-21.1.8} compiler, 
and the Rust benchmarks were compiled using \texttt{rustc-1.94.1}. Both \texttt{clang-21.1.8} and \texttt{rustc-1.94.1}
use the \texttt{LLVM-21} backend to compile to \texttt{x86\_64} assembly code. This ensures that
both the C and Rust benchmarks are optimized by the LLVM optimization pipeline.
To isolate the impact of Rust's memory safety checks, we use a modified version of Rust that disables memory-safety related runtime checks~\cite{zhang2022towards}, which we call the \rmm{}
compiler. 
In particular, the following safety checks were disabled in this compiler version---bounds checking for arrays, 
vectors, slices, and string/byte slices, and integer overflow detection, which
is disabled in the base Rust version as well.
All other checks, such as reference counting, remain intact.
All benchmarks were evaluated at the \texttt{-O3} optimization mode,
using Release Mode builds.

\para{Perf and Pin tools.}
To characterize microarchitectural behavior, we use hardware performance counters~\cite{perf, de2010new, perf-event-open-man} to collect branch mispredictions, L1-D cache misses, and L2 cache misses and instruction-level instrumentation with Intel Pin~\cite{pin, pintool} to obtain dynamic traces of all executed branches, loads, and stores. We do not report L3 misses because the L3 cache is shared across cores and can be interrupted by OS activity \cite{fedorova2010managing}. To the best of our knowledge per-process or per-core metrics are not provided by either vendor. 
We focus our analysis on cache and branch-related behavior, as they are the primary microarchitectural components that account for the performance differences observed between Rust and C. 

\emph{AI Usage.} The scripts that produce the plots in this paper were generated with
Claude Code.
All plotted data was collected from our own experiments
using the tooling described above.

\section{Performance Characterization Summary}
\label{summary}

\subsection{Performance Across C and Rust}

\begin{figure}[t]
    \centering
    \includegraphics[width=\textwidth]{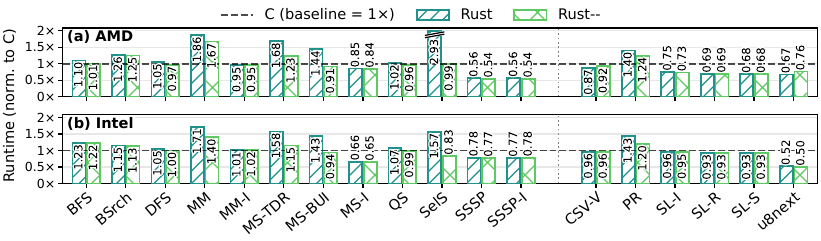}
    \caption{Runtime of Rust and \rmm{} normalized to C on (a) AMD and (b) Intel
    ($1\times$ = C runtime; lower is better). Bars above $2\times$ are truncated.
    The dotted line separates microbenchmarks (left) from macrobenchmarks (right).
    Benchmark abbreviations:
BSrch (Binary Search), MM (Matrix Multiplication), MS (Merge Sort),
QS (Quicksort), SelS (Selection Sort), SL (Skiplist),
PR (PageRank), CSV-V (CSV Validation), and I (Iterator).
TDR and BUI denote top-down recursive and bottom-up iterative
implementations, respectively.}
    \label{fig:runtime_overhead}
\end{figure}

\autoref{fig:runtime_overhead} shows the runtime overhead of the benchmarks for AMD and Intel machines, respectively.
To minimize variance, all reported results are averaged across three independent runs.
We observe that despite the Rust code performing additional memory-safety checks,
for 8 out of 12 benchmarks, at least one implementation variant incurs less than 10\%
overhead over the C implementation, and 5 \emph{outperformed} the C implementation on both architectures.
Moreover, for Rust, the iterator-based Matrix Multiplication and Merge Sort implementations, the UTF-8 Decode (\texttt{u8next})
benchmark, the LibCsv, and the Skip List benchmarks \emph{outperformed} their C counterparts.
Even among the index-based implementations that use random access into vectors---a pattern that necessarily incurs bounds 
checks on every access---several benchmarks exhibit modest overhead. 
This demonstrates that bounds checks alone do not necessarily lead to significant performance degradation.

\definecolor{colA}{gray}{1.00}
\definecolor{colB}{gray}{0.94}
\definecolor{myred}{RGB}{200,30,30}
\definecolor{mygreen}{RGB}{0,150,0}

\definecolor{colA}{gray}{1.00}
\definecolor{colB}{gray}{0.94}
\definecolor{myred}{RGB}{200,30,30}
\definecolor{mygreen}{RGB}{0,150,0}

\subsection{Safety Checks Impact.} We observe that runtime 
bounds checks contribute to
performance overhead for 
several, but not all, benchmarks.
By using the \rmm{} 
compiler, we isolate the effect of runtime checks on performance. Eliminating bounds checks substantially reduces overhead for Matrix Multiplication (from 1.86$\times$ to 1.67$\times$ on AMD, 1.71$\times$ to 1.40$\times$ on Intel), recursive Merge Sort with indexing (from 1.68$\times$ to 1.23$\times$ on AMD, 1.58$\times$ to 1.15$\times$ on Intel), and PageRank (from 1.40$\times$ to 1.24$\times$ on AMD, 1.43$\times$ to 1.20$\times$ on Intel), but does not account for all of it. However, for Selection Sort
and iterative Merge Sort, bounds check elimination removes the overhead entirely. Selection Sort drops from 2.93$\times$ to 0.99$\times$ on AMD and from 1.57$\times$ to 0.83$\times$ on Intel, while iterative Merge Sort drops from 1.44$\times$ to 0.91$\times$ on AMD and from 1.43$\times$ to 0.94$\times$ on Intel. In both of these cases, performance matches or outperforms C. In contrast, bounds check elimination makes less than a 10\% difference on BFS, DFS, and Binary Search, despite their presence (as suggested by the decrease in conditional branches ranging from 30\% to 60\% comparing to \rmm{}). In these cases, Rust's performance overhead originates elsewhere. We investigate these overhead sources in \autoref{sec:microeffects}. 

Moreover, out of the total 18 benchmarks and variants in our dataset, even the \rmm{} version
underperformed the baseline C version for 4 (AMD) and 5 (Intel) benchmarks, respectively. Out of 
these BFS and Binary Search
exhibited more than 20\% overhead than the C version. This shows that
there exists overhead sources in addition to Rust's runtime memory safety checks.
Previous
work by Zhang et al.~\cite{zhang2022towards}
attributed any remaining performance overhead after disabling the 
memory-safety checks to ASCII to UTF-8 conversion for string-based workloads. 
However, our Rust benchmarks do 
not perform any additional ASCII to UTF-8 conversion over the C baseline. 
As we will discuss in detail in \autoref{sec:microeffects} 
and \autoref{sec:compilereffects},
the additional overhead occurs due to a complex interplay between compiler
optimizations and microarchitectural components. 
Thus, the performance of
Rust is not just the cost of the safety checks over the C baseline runtime.

\subsection{Implementation Strategy Impact}
\label{sec:programming_idiom}
We observed that the implementation strategy played a significant 
role in the Rust overhead, and also 
that in terms of performance, no particular
strategy consistently outperformed others.

The Merge Sort algorithm has two algorithmic
variants---a top-down recursive (TDR), and a bottom-up iterative (BUI) variant. The top-down recursive (TDR) variant follows the classic divide-and-conquer approach, recursively splitting the array in half until reaching single-element subarrays, then merging upward. Each call to \texttt{merge} allocates and frees two temporary arrays for the left and right halves. The bottom-up iterative (BUI) variant instead uses a single pre-allocated temporary buffer and iteratively merges subarrays of increasing width (1, 2, 4, …), copying the full result back after each pass.
Because the bottom-up variant skips making additional memory copies, one
would expect that it outperforms the top-down variant. We indeed observe 
this effect, with the C top-down benchmark taking $\sim$12 seconds
to sort a 100-million
sized integer array, while the bottom-up benchmark
takes $\sim$9.5 seconds on both AMD and Intel architectures. One might expect that the overhead of
the Rust translations of both these algorithm variants would remain
consistent. However, as shown in \autoref{fig:runtime_overhead}, 
Rust has a 68\% overhead for the top-down algorithm, but 
a 44\% overhead for the bottom-up algorithm on the AMD, and 58\% and 43\%, for Intel, respectively.
Thus, not only is the top-down variant \emph{slower for the baseline C}, it has a \emph{higher overhead when transpiled to Rust}.
We discuss potential 
reasons for this behavior using microarchitectural metrics in 
\autoref{sec:hiding_bp}.

Additionally, it is possible to implement the 
same algorithm in both imperative and functional programming styles.
For example, the sort algorithms can be implemented in an imperative-style,
using indexing operations
into the \texttt{vec} objects, or using a more functional programming style,
using iterators and \texttt{map} and \texttt{fold} operations. 
One might expect that because the iterator-based implementations
do not involve runtime bounds checks, they would \emph{always} outperform the 
indexing-based implementations. However, as shown in \autoref{fig:runtime_overhead}
we observed that this was
true only for Matrix Multiplication and
Merge Sort, where the iterator-style 
implementations matched or \emph{improved} the Rust runtime over C. 
For SSSP, the iterator and indexed implementations achieved nearly identical performance, with both outperforming the C implementation. In contrast, Quicksort suffered little indexing overhead, with performance comparable to C.  
This shows that despite programmer intuition, 
there is a lack of a definitive superior implementation
strategy.

\section{\sysname: An AI-assisted Performance Debugging Framework}
\label{sec:ferroscope}

\begin{figure}[t]
  \centering
  \includegraphics[width=\textwidth]{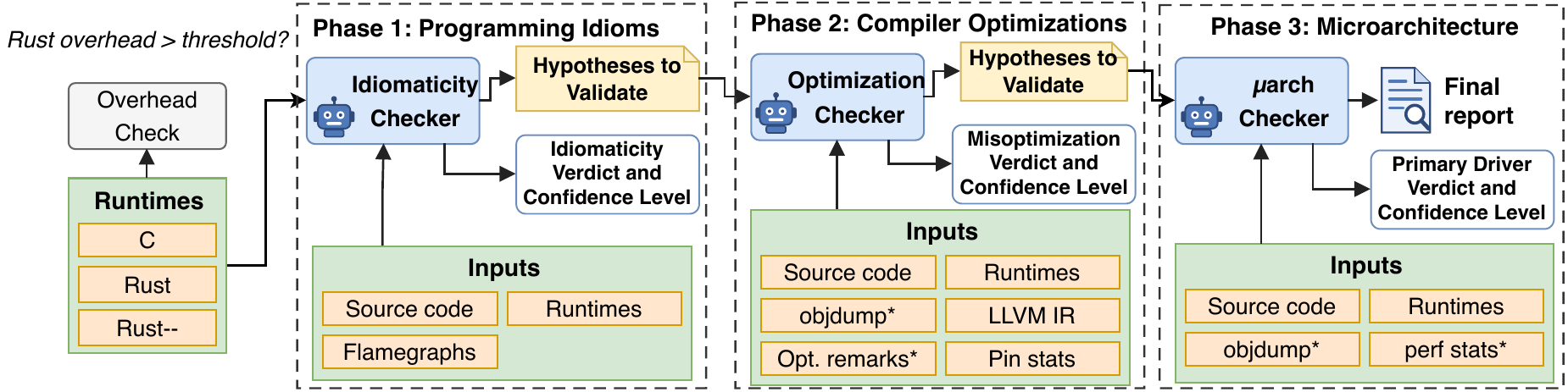}
  \caption{\sysname's AI-assisted performance debugging pipeline. The phases run only when the Rust runtime exceeds the C runtime by more than the overhead threshold (10\% in our evaluation). Each phase records a verdict and a confidence level (low, medium, or high) in the final report. Phases~1 and 2 answer yes or no to their question, and Phase~3 names the primary driver of the overhead. Only the notes pass to the next phase, as leads that it must verify in its own inputs. Inputs marked * also include the \rmm{} build.}
  \label{fig:ai-debug-pipeline}
\end{figure}

We developed \sysname to harness the capabilities of AI to systematically
identify and explain the root causes of performance differences
between C programs and their Rust counterparts. 
As illustrated in \autoref{fig:ai-debug-pipeline}, \sysname
consists of three sequential phases which
consume 
debugging and profiling
information relevant to cross-layer performance analysis, including
call-stack profiles (flamegraphs), compiler
intermediate representations, optimization remarks, dynamic
instruction traces, and hardware performance counters.
The pipeline is triggered only when the Rust implementation's
execution time exceeds that of the corresponding C
implementation by a configurable threshold (10\% in our
evaluation). Each phase investigates a distinct potential
source of performance degradation, namely,
programming idioms,
compiler optimizations, or microarchitectural effects.
Phases 1 and 2 produce a user-facing verdict describing whether
the investigated factor contributes to the observed overhead,
along with a confidence level (low, medium, or high) reflecting
the strength of the supporting evidence. The final phase
identifies the primary microarchitectural driver
of the overhead. Each phase also generates hypotheses for
subsequent phases to validate, supported by observations
from its own inputs. These hypotheses are passed forward
as diagnostic notes, allowing later phases to build on
earlier findings without assuming their correctness.
A subsequent phase must independently validate a hypothesis
using its own evidence before incorporating it into its
conclusions. Only these notes are propagated between phases;
the verdicts, confidence levels, and summaries are collected
into the final diagnostic report.
We instantiate the pipeline with Claude Opus~5.5 in all three
phases, although the pipeline itself is model-agnostic. 

\para{Phase 1: Programming Idioms.}
Motivated by our findings on the performance implications of
programming idioms in~\autoref{sec:programming_idiom},
the first phase investigates whether differences in
programming idioms or implementation strategies between
C and Rust contribute to the observed performance degradation.
It analyzes the source code of both implementations,
their execution times, and call-stack profiles
(flamegraphs) to identify potentially expensive Rust
abstractions, unnecessary computations, or differences
in algorithmic implementation. The phase reports a
positive verdict only when an implementation-level
difference in performance-relevant code plausibly
explains a substantial portion of the observed overhead.
It also identifies potential compiler-level consequences
of these implementation choices for further investigation.

\para{Phase 2: Compiler Optimizations.}
Phase 2 investigates whether differences in
compiler-generated code contribute to the observed
performance degradation. It analyzes the source code,
LLVM intermediate representations, compiler optimization
remarks, disassembly, and dynamic instruction counts
collected using Pin tools. By comparing the C and Rust
implementations across these representations, the phase
identifies missed optimizations, redundant runtime checks,
differences in instruction selection, and other code
generation inefficiencies. A positive verdict 
is reported only when there is
evidence of a concrete optimization opportunity or
inefficiency in the generated code that is not required
by the program's semantics. The phase also generates
hypotheses about the performance consequences of these
differences for further microarchitectural analysis.

\para{Phase 3: Microarchitectural Effects.}
The final phase investigates how the generated instructions
interact with the underlying hardware to explain the
observed runtime differences. It combines the
disassembled assembly instructions,
dynamic instruction statistics, and hardware performance
counters, including branch mis-predictions and cache misses,
to distinguish overhead arising from additional instructions,
reduced instruction-level parallelism, memory behavior,
and other microarchitectural effects. This phase identifies the primary driver
of the performance difference and assigns a confidence
level based on the available evidence. Its findings also
help validate or refine hypotheses generated by the
preceding phases.

\emph{Isolating Runtime Safety Costs.}
Where applicable, the pipeline additionally uses measurements
from \rmm{}, a modified Rust compiler that disables selected
runtime safety checks, to distinguish their costs from other
performance effects. All three phases receive the \rmm{}
execution time, while Phase~2 additionally receives its
disassembly and optimization remarks, and Phase~3 receives
its disassembly and hardware performance counters.
Comparing standard Rust against \rmm{} helps identify
whether the overhead originates from runtime checks
themselves or from other compiler and microarchitectural
effects.

\section{Understanding Performance Differences}
\autoref{tab:bench_perf_summary} shows the high-level
summary of the analysis results. These results
draw on our manual cross-layer analysis and are
corroborated by \sysname's AI-assisted diagnosis.
In several cases, \sysname further refines these
explanations by identifying additional contributing
factors or specific compiler-level causes. 
\autoref{sec:microeffects} and \autoref{sec:compilereffects} presents detailed
results on how micro-architectural optimizations
and compiler effects impact the translated Rust
code's performance. \autoref{tab:ratios_intel} and \autoref{tab:ratios_amd} show the dynamic instruction counts and performance counters of Rust and \rmm{} relative to C. We note the difference in dynamic instruction mix between the two architectures, which we explain in \autoref{app:inst-mix-diff}.

\begin{table*}
\centering
\caption{
Summary: Benchmark Variants and Rust's Relative Performance to C at \texttt{-O3} The benchmarks with * indicate cases
where the \sysname analysis found additional insights 
over those reported by manual analysis. 
}
\small
\resizebox{\textwidth}{!}{
\begin{tabular}{L{0.22\linewidth} L{0.15\linewidth} L{0.08\linewidth} L{0.08\linewidth} L{0.47\linewidth}}
\toprule
\textbf{Benchmark} & \textbf{Variant} & \multicolumn{2}{c}{\textbf{Overhead}} & \multicolumn{1}{c}{\textbf{Explanation}} \\
\cmidrule(lr){3-4}
& & \textbf{AMD} & \textbf{Intel} & \\
\midrule
Binary Search & \texttt{index-based} & $\uparrow$ 26\% & $\uparrow$ 15\% & \raggedcentercell{LLVM lowers branches to \texttt{CMOV}s, serializing consecutive dependent loads.}\\
\cmidrule(lr){1-5}
Depth-First Search & \texttt{index-based} & $\uparrow$ 5\% & $\uparrow$ 5\% & \raggedcentercell{Lower L1D and L2D misses offset the increase in memory and branch instructions.}\\
\cmidrule(lr){1-5}
Breadth-First Search & \texttt{index-based} & $\uparrow$ 10\% & $\uparrow$ 23\% & \raggedcentercell{Lost alias information prevents hoisting row pointers, introducing dependent loads before unpredictable branches.*}\\
\cmidrule(lr){1-5}
\multirow{2}{*}{\shortstack[l]{Single-Source\\Shortest Path (SSSP)}} &
\texttt{index-based} & $\downarrow$ 44\% & $\downarrow$ 22\% & \multirow{2}{*}{\raggedcentercell{C uses \texttt{CMOV}s and stores every min computation, while Rust only stores when an update is needed.}}\\
\cmidrule(lr){2-4}
& \texttt{iterator} & $\downarrow$ 44\% & $\downarrow$ 23\% & \\
\cmidrule(lr){1-5}
Selection Sort & \texttt{index-based} & $\uparrow$ 193\% & $\uparrow$ 57\% & \raggedcentercell{A redundant bounds check prevents 4$\times$ loop unrolling, substantially increasing instructions per element. AMD also suffers from poor instruction layout.*}\\
\cmidrule(lr){1-5}
Quicksort & \texttt{index-based} & $\uparrow$ 2\% & $\uparrow$ 7\% & \raggedcentercell{High branch predictor accuracy mitigates increased bounds checks with help from matching or better cache performance.}\\
\cmidrule(lr){1-5}
\multirow{3}{*}{Merge Sort} &
\texttt{index-based (TDR)} & $\uparrow$ 68\% & $\uparrow$ 58\% & \raggedcentercell{Bounds checks prevent \texttt{CMOV} conversion, introducing unpredictable branches. Zero-initialized temporary buffers add allocation overhead.*}\\
\cmidrule(lr){2-5}
& \texttt{iterator} & $\downarrow$ 15\% & $\downarrow$ 34\% & \raggedcentercell{Eliminates bounds checks and unpredictable branches leading to fewer mispredictions.}\\
\cmidrule(lr){2-5}
& \texttt{index-based (BUI)} & $\uparrow$ 44\% & $\uparrow$ 43\% & \raggedcentercell{Bounds checks with distinct panic locations prevent \texttt{CMOV} conversion, resulting in unpredictable branches.}\\
\cmidrule(lr){1-5}
\multirow{2}{*}{\shortstack[l]{Matrix\\Multiplication}} &
\texttt{index-based} & $\uparrow$ 86\% & $\uparrow$ 71\% & \raggedcentercell{Lost alias information prevents loop-invariant loads and register accumulation, introducing repeated loads and a loop-carried dependency.*}\\
\cmidrule(lr){2-5}
& \texttt{iterator} & $\downarrow$ 5\% & $\uparrow$ 1\% & \raggedcentercell{Eliminated bounds checks leads to same performance as C.}\\
\cmidrule(lr){1-5}
PageRank &
\shortstack{\texttt{index-based}} & $\uparrow$ 40\% & $\uparrow$ 43\% & \raggedcentercell{Escaped variables prevent elimination of some redundant checks. Additional memory traffic from 8-byte indices is a possible contributor, but the remaining overhead is unexplained.*}\\
\cmidrule(lr){1-5}
UTF-8 Decode (\texttt{u8next}) & \texttt{index-based} & $\downarrow$ 33\% & $\downarrow$ 48\% & \raggedcentercell{Cache misses nearly eliminated (Intel) or greatly reduced while retaining similar instruction count to C.}\\
\cmidrule(lr){1-5}
CSV Validation (LibCSV) & \texttt{index-based} & $\downarrow$ 13\% & $\downarrow$ 4\% & \raggedcentercell{Decrease in instruction count and improved cache performance.}\\
\cmidrule(lr){1-5}
Skip List & \texttt{index-based} & avg. $\downarrow$ 29.3\% & avg. $\downarrow$ 6\% & \raggedcentercell{Lower L1D and L2D misses outweighs increased dynamic memory instructions. High branch prediction accuracy mitigates additional conditional branches.}\\
\bottomrule
\end{tabular}
}
\label{tab:bench_perf_summary}
\end{table*}

\subsection{Micro-architectural Optimizations Hiding Rust Overheads}
\label{sec:microeffects}

\begin{table*}
\centering
\setlength{\tabcolsep}{4pt}
\resizebox{\textwidth}{!}{%
\begin{tabular}{@{}lrrrrrrrrrrrrrrrrrrrr@{}}
\toprule
\textbf{App.} & \multicolumn{2}{c}{\textbf{Time}} & \multicolumn{9}{c}{\textbf{Rust}} & \multicolumn{9}{c}{\textbf{\rmm{}}} \\
\cmidrule(lr){2-3} \cmidrule(lr){4-12} \cmidrule(lr){13-21}
 & Rust & \rmm{} & \textbf{IPC} & \textbf{Cycles} & \makecell[r]{\textbf{Total} \\ \textbf{Inst.}} & \makecell[r]{\textbf{Cond} \\ \textbf{Br}} & \textbf{Stores} & \textbf{Loads} & \makecell[r]{\textbf{Br} \\ \textbf{Misses}} & \makecell[r]{\textbf{L1D} \\ \textbf{Misses}} & \makecell[r]{\textbf{L2D} \\ \textbf{Misses}} & \textbf{IPC} & \textbf{Cycles} & \textbf{Inst.} & \makecell[r]{\textbf{Cond} \\ \textbf{Br}} & \textbf{Stores} & \textbf{Loads} & \makecell[r]{\textbf{Br} \\ \textbf{Misses}} & \makecell[r]{\textbf{L1D} \\ \textbf{Misses}} & \makecell[r]{\textbf{L2D} \\ \textbf{Misses}} \\
\midrule
\rowcolor{colA} \textbf{BFS} & \textcolor{myred}{1.23$\times$} & \textcolor{myred}{1.22$\times$} & 1.40$\times$ & 1.19$\times$ & 1.67$\times$ & 1.40$\times$ & 2.73$\times$ & 3.00$\times$ & 1.00$\times$ & 1.24$\times$ & 0.81$\times$ & 0.96$\times$ & 1.22$\times$ & 1.17$\times$ & 1.00$\times$ & 2.73$\times$ & 2.33$\times$ & 1.00$\times$ & 1.17$\times$ & 0.80$\times$ \\
\rowcolor{colB} \textbf{BSrch} & \textcolor{myred}{1.15$\times$} & \textcolor{myred}{1.13$\times$} & 1.27$\times$ & 1.16$\times$ & 1.48$\times$ & 1.04$\times$ & 1.11$\times$ & 1.02$\times$ & 0.06$\times$ & 0.35$\times$ & 0.28$\times$ & 1.20$\times$ & 1.13$\times$ & 1.35$\times$ & 0.74$\times$ & 1.11$\times$ & 1.02$\times$ & 0.06$\times$ & 0.35$\times$ & 0.28$\times$ \\
\rowcolor{colA} \textbf{DFS} & 1.05$\times$ & \textcolor{mygreen}{1.00$\times$} & 1.41$\times$ & 1.12$\times$ & 1.57$\times$ & 1.60$\times$ & 1.43$\times$ & 1.40$\times$ & 0.99$\times$ & 0.51$\times$ & 0.76$\times$ & 1.00$\times$ & 1.00$\times$ & 1.00$\times$ & 1.00$\times$ & 1.29$\times$ & 1.00$\times$ & 0.99$\times$ & 0.52$\times$ & 0.92$\times$ \\
\rowcolor{colB} \textbf{MM} & \textcolor{myred}{1.71$\times$} & \textcolor{myred}{1.40$\times$} & 2.02$\times$ & 1.60$\times$ & 3.23$\times$ & 9.97$\times$ & 1.00$\times$ & 3.00$\times$ & 1.00$\times$ & 1.07$\times$ & 1.01$\times$ & 0.91$\times$ & 1.35$\times$ & 1.23$\times$ & 1.00$\times$ & 1.00$\times$ & 2.00$\times$ & 1.00$\times$ & 1.01$\times$ & 1.01$\times$ \\
\rowcolor{colA} \textbf{MM-I} & 1.01$\times$ & 1.02$\times$ & 0.99$\times$ & 1.01$\times$ & 1.00$\times$ & 1.01$\times$ & 0.67$\times$ & 1.00$\times$ & 1.00$\times$ & 1.96$\times$ & 1.53$\times$ & 1.00$\times$ & 1.00$\times$ & 1.00$\times$ & 1.00$\times$ & 0.67$\times$ & 1.00$\times$ & 1.00$\times$ & 2.13$\times$ & 1.59$\times$ \\
\rowcolor{colB} \textbf{MS-TDR} & \textcolor{myred}{1.58$\times$} & \textcolor{myred}{1.15$\times$} & 0.79$\times$ & 1.49$\times$ & 1.18$\times$ & 2.13$\times$ & 1.46$\times$ & 1.26$\times$ & 18.19$\times$ & 1.07$\times$ & 1.72$\times$ & 1.11$\times$ & 1.14$\times$ & 1.26$\times$ & 1.61$\times$ & 1.76$\times$ & 1.60$\times$ & 0.98$\times$ & 0.96$\times$ & 0.80$\times$ \\
\rowcolor{colA} \textbf{MS-BUI} & \textcolor{myred}{1.43$\times$} & \textcolor{mygreen}{0.94$\times$} & 0.66$\times$ & 1.41$\times$ & 0.93$\times$ & 2.69$\times$ & 1.10$\times$ & 1.00$\times$ & 20.99$\times$ & 0.94$\times$ & 1.46$\times$ & 1.08$\times$ & 0.92$\times$ & 0.99$\times$ & 0.98$\times$ & 1.09$\times$ & 1.46$\times$ & 1.00$\times$ & 0.95$\times$ & 0.97$\times$ \\
\rowcolor{colB} \textbf{MS-I} & \textcolor{mygreen}{0.66$\times$} & \textcolor{mygreen}{0.65$\times$} & 1.72$\times$ & 0.63$\times$ & 1.08$\times$ & 1.53$\times$ & 1.65$\times$ & 1.34$\times$ & 0.05$\times$ & 0.97$\times$ & 0.75$\times$ & 1.68$\times$ & 0.64$\times$ & 1.08$\times$ & 1.53$\times$ & 1.65$\times$ & 1.34$\times$ & 0.05$\times$ & 0.97$\times$ & 0.76$\times$ \\
\rowcolor{colA} \textbf{QS} & 1.07$\times$ & \textcolor{mygreen}{0.99$\times$} & 1.21$\times$ & 1.07$\times$ & 1.29$\times$ & 1.63$\times$ & 1.04$\times$ & 1.02$\times$ & 1.03$\times$ & 0.59$\times$ & 0.87$\times$ & 0.95$\times$ & 0.99$\times$ & 0.94$\times$ & 1.01$\times$ & 0.97$\times$ & 0.98$\times$ & 0.99$\times$ & 1.23$\times$ & 1.01$\times$ \\
\rowcolor{colA} \textbf{SelS} & \textcolor{myred}{1.57$\times$} & \textcolor{mygreen}{0.83$\times$} & 1.36$\times$ & 1.60$\times$ & 2.18$\times$ & 2.40$\times$ & 1.00$\times$ & 1.00$\times$ & 1.00$\times$ & 3.03$\times$ & 2.97$\times$ & 0.98$\times$ & 0.84$\times$ & 0.82$\times$ & 1.00$\times$ & 1.00$\times$ & 1.00$\times$ & 1.03$\times$ & 2.82$\times$ & 32.70$\times$ \\
\rowcolor{colB} \textbf{SSSP} & \textcolor{mygreen}{0.78$\times$} & \textcolor{mygreen}{0.77$\times$} & 1.25$\times$ & 0.75$\times$ & 0.94$\times$ & 2.50$\times$ & 0.00$\times$ & 1.00$\times$ & 1758.61$\times$ & 1.44$\times$ & 2.25$\times$ & 0.83$\times$ & 0.75$\times$ & 0.63$\times$ & 1.25$\times$ & 0.00$\times$ & 1.00$\times$ & 1767.93$\times$ & 0.26$\times$ & 3.26$\times$ \\
\rowcolor{colA} \textbf{SSSP-I} & \textcolor{mygreen}{0.77$\times$} & \textcolor{mygreen}{0.78$\times$} & 1.24$\times$ & 0.76$\times$ & 0.94$\times$ & 2.50$\times$ & 0.00$\times$ & 1.00$\times$ & 1772.72$\times$ & 1.44$\times$ & 2.45$\times$ & 0.91$\times$ & 0.75$\times$ & 0.69$\times$ & 1.50$\times$ & 0.00$\times$ & 1.00$\times$ & 1790.17$\times$ & 1.50$\times$ & 2.70$\times$ \\
\midrule
\rowcolor{colB} \textbf{CSV-V} & \textcolor{mygreen}{0.96$\times$} & \textcolor{mygreen}{0.96$\times$} & 0.78$\times$ & 0.93$\times$ & 0.73$\times$ & 0.91$\times$ & 0.66$\times$ & 0.67$\times$ & 0.78$\times$ & 0.13$\times$ & 0.34$\times$ & 0.78$\times$ & 0.93$\times$ & 0.73$\times$ & 0.91$\times$ & 0.66$\times$ & 0.67$\times$ & 0.78$\times$ & 0.19$\times$ & 0.57$\times$ \\
\rowcolor{colA} \textbf{PR} & \textcolor{myred}{1.43$\times$} & \textcolor{myred}{1.20$\times$} & 1.04$\times$ & 1.45$\times$ & 1.50$\times$ & 2.11$\times$ & 0.74$\times$ & 2.01$\times$ & 0.96$\times$ & 1.06$\times$ & 1.17$\times$ & 0.80$\times$ & 1.20$\times$ & 0.96$\times$ & 1.02$\times$ & 1.03$\times$ & 1.49$\times$ & 0.97$\times$ & 1.04$\times$ & 1.11$\times$ \\
\rowcolor{colB} \textbf{SL-I} & \textcolor{mygreen}{0.96$\times$} & \textcolor{mygreen}{0.95$\times$} & 2.66$\times$ & 0.94$\times$ & 2.51$\times$ & 3.37$\times$ & 3.80$\times$ & 2.47$\times$ & 0.99$\times$ & 0.63$\times$ & 0.76$\times$ & 2.49$\times$ & 0.93$\times$ & 2.31$\times$ & 2.86$\times$ & 3.71$\times$ & 2.14$\times$ & 1.01$\times$ & 0.64$\times$ & 0.78$\times$ \\
\rowcolor{colA} \textbf{SL-R} & \textcolor{mygreen}{0.93$\times$} & \textcolor{mygreen}{0.93$\times$} & 2.64$\times$ & 0.93$\times$ & 2.45$\times$ & 3.51$\times$ & 4.58$\times$ & 2.06$\times$ & 0.98$\times$ & 0.63$\times$ & 0.77$\times$ & 2.40$\times$ & 0.92$\times$ & 2.21$\times$ & 3.03$\times$ & 4.42$\times$ & 1.92$\times$ & 0.99$\times$ & 0.63$\times$ & 0.77$\times$ \\
\rowcolor{colB} \textbf{SL-S} & \textcolor{mygreen}{0.93$\times$} & \textcolor{mygreen}{0.93$\times$} & 1.98$\times$ & 0.94$\times$ & 1.86$\times$ & 2.71$\times$ & 6.92$\times$ & 1.78$\times$ & 1.00$\times$ & 0.64$\times$ & 0.79$\times$ & 1.87$\times$ & 0.92$\times$ & 1.73$\times$ & 2.35$\times$ & 6.92$\times$ & 1.64$\times$ & 0.96$\times$ & 0.65$\times$ & 0.80$\times$ \\
\rowcolor{colA} \textbf{u8next} & \textcolor{mygreen}{0.52$\times$} & \textcolor{mygreen}{0.50$\times$} & 2.14$\times$ & 0.51$\times$ & 1.09$\times$ & 1.12$\times$ & 1.00$\times$ & 1.00$\times$ & 1.20$\times$ & 0.03$\times$ & 0.00$\times$ & 2.20$\times$ & 0.50$\times$ & 1.09$\times$ & 1.12$\times$ & 1.00$\times$ & 1.00$\times$ & 0.93$\times$ & 0.03$\times$ & 0.00$\times$ \\
\bottomrule
\end{tabular}%
}
\caption{Rust/C performance ratios (Intel). 
Benchmark
abbreviations follow~\autoref{tab:bench_perf_summary}.
Time values $\leq 1.00\times$ are in \textcolor{mygreen}{green}
and values $\geq 1.10\times$ are in \textcolor{myred}{red}.}
\label{tab:ratios_intel}
\end{table*}

\begin{table*}
\centering
\setlength{\tabcolsep}{4pt}
\resizebox{\textwidth}{!}{%
\begin{tabular}{@{}lrrrrrrrrrrrrrrrrrrrr@{}}
\toprule
\textbf{App.} & \multicolumn{2}{c}{\textbf{Time}} & \multicolumn{9}{c}{\textbf{Rust}} & \multicolumn{9}{c}{\textbf{\rmm{}}} \\
\cmidrule(lr){2-3} \cmidrule(lr){4-12} \cmidrule(lr){13-21}
 & Rust & \rmm{} & \textbf{IPC} & \textbf{Cycles} & \makecell[r]{\textbf{Total} \\ \textbf{Inst.}} & \makecell[r]{\textbf{Cond} \\ \textbf{Br}} & \textbf{Stores} & \textbf{Loads} & \makecell[r]{\textbf{Br} \\ \textbf{Misses}} & \makecell[r]{\textbf{L1D} \\ \textbf{Misses}} & \makecell[r]{\textbf{L2D} \\ \textbf{Misses}} & \textbf{IPC} & \textbf{Cycles} & \textbf{Inst.} & \makecell[r]{\textbf{Cond} \\ \textbf{Br}} & \textbf{Stores} & \textbf{Loads} & \makecell[r]{\textbf{Br} \\ \textbf{Misses}} & \makecell[r]{\textbf{L1D} \\ \textbf{Misses}} & \makecell[r]{\textbf{L2D} \\ \textbf{Misses}} \\
\midrule
\rowcolor{colA} \textbf{BFS} & \textcolor{myred}{1.10$\times$} & 1.01$\times$ & 1.37$\times$ & 1.22$\times$ & 1.67$\times$ & 1.40$\times$ & 2.51$\times$ & 3.00$\times$ & 1.00$\times$ & 1.01$\times$ & 1.01$\times$ & 1.04$\times$ & 1.12$\times$ & 1.17$\times$ & 1.00$\times$ & 2.51$\times$ & 2.33$\times$ & 1.00$\times$ & 1.02$\times$ & 1.01$\times$ \\
\rowcolor{colB} \textbf{BSrch} & \textcolor{myred}{1.26$\times$} & \textcolor{myred}{1.25$\times$} & 1.16$\times$ & 1.27$\times$ & 1.48$\times$ & 1.04$\times$ & 1.11$\times$ & 1.02$\times$ & 0.06$\times$ & 0.27$\times$ & 0.27$\times$ & 1.07$\times$ & 1.26$\times$ & 1.35$\times$ & 0.74$\times$ & 1.11$\times$ & 1.02$\times$ & 0.06$\times$ & 0.27$\times$ & 0.27$\times$ \\
\rowcolor{colA} \textbf{DFS} & 1.05$\times$ & \textcolor{mygreen}{0.97$\times$} & 1.60$\times$ & 0.98$\times$ & 1.57$\times$ & 1.60$\times$ & 1.43$\times$ & 1.40$\times$ & 1.00$\times$ & 0.63$\times$ & 0.95$\times$ & 1.03$\times$ & 0.97$\times$ & 1.00$\times$ & 1.00$\times$ & 1.29$\times$ & 1.00$\times$ & 1.00$\times$ & 0.63$\times$ & 0.97$\times$ \\
\rowcolor{colB} \textbf{MM} & \textcolor{myred}{1.86$\times$} & \textcolor{myred}{1.67$\times$} & 1.75$\times$ & 1.84$\times$ & 3.23$\times$ & 9.97$\times$ & 1.00$\times$ & 3.00$\times$ & 1.00$\times$ & 1.22$\times$ & 1.20$\times$ & 0.74$\times$ & 1.66$\times$ & 1.23$\times$ & 1.00$\times$ & 1.00$\times$ & 2.00$\times$ & 1.00$\times$ & 1.21$\times$ & 1.13$\times$ \\
\rowcolor{colA} \textbf{MM-I} & \textcolor{mygreen}{0.95$\times$} & \textcolor{mygreen}{0.95$\times$} & 1.06$\times$ & 0.95$\times$ & 1.00$\times$ & 1.01$\times$ & 0.67$\times$ & 1.00$\times$ & 1.00$\times$ & 1.01$\times$ & 1.03$\times$ & 1.06$\times$ & 0.94$\times$ & 1.00$\times$ & 1.00$\times$ & 0.67$\times$ & 1.00$\times$ & 1.00$\times$ & 1.01$\times$ & 1.03$\times$ \\
\rowcolor{colB} \textbf{MS-TDR} & \textcolor{myred}{1.68$\times$} & \textcolor{myred}{1.23$\times$} & 0.70$\times$ & 1.69$\times$ & 1.18$\times$ & 2.13$\times$ & 1.23$\times$ & 1.39$\times$ & 18.60$\times$ & 1.21$\times$ & 1.16$\times$ & 1.03$\times$ & 1.23$\times$ & 1.27$\times$ & 1.62$\times$ & 1.25$\times$ & 1.56$\times$ & 1.01$\times$ & 1.17$\times$ & 2.09$\times$ \\
\rowcolor{colA} \textbf{MS-BUI} & \textcolor{myred}{1.44$\times$} & \textcolor{mygreen}{0.91$\times$} & 0.62$\times$ & 1.51$\times$ & 0.93$\times$ & 2.69$\times$ & 1.10$\times$ & 1.00$\times$ & 20.52$\times$ & 1.00$\times$ & 1.00$\times$ & 1.13$\times$ & 0.88$\times$ & 0.99$\times$ & 0.98$\times$ & 1.09$\times$ & 1.46$\times$ & 1.00$\times$ & 1.00$\times$ & 0.95$\times$ \\
\rowcolor{colB} \textbf{MS-I} & \textcolor{mygreen}{0.85$\times$} & \textcolor{mygreen}{0.84$\times$} & 1.36$\times$ & 0.80$\times$ & 1.09$\times$ & 1.53$\times$ & 1.12$\times$ & 1.27$\times$ & 0.09$\times$ & 1.04$\times$ & 2.02$\times$ & 1.37$\times$ & 0.79$\times$ & 1.09$\times$ & 1.53$\times$ & 1.12$\times$ & 1.27$\times$ & 0.09$\times$ & 1.04$\times$ & 2.00$\times$ \\
\rowcolor{colA} \textbf{QS} & 1.02$\times$ & \textcolor{mygreen}{0.96$\times$} & 1.29$\times$ & 1.00$\times$ & 1.29$\times$ & 1.63$\times$ & 1.04$\times$ & 1.02$\times$ & 0.98$\times$ & 1.00$\times$ & 0.95$\times$ & 0.98$\times$ & 0.96$\times$ & 0.94$\times$ & 1.01$\times$ & 0.97$\times$ & 0.98$\times$ & 0.99$\times$ & 1.00$\times$ & 1.00$\times$ \\
\rowcolor{colA} \textbf{SelS} & \textcolor{myred}{2.93$\times$} & \textcolor{mygreen}{0.99$\times$} & 0.75$\times$ & 2.90$\times$ & 2.18$\times$ & 2.40$\times$ & 1.00$\times$ & 1.00$\times$ & 0.94$\times$ & 1.00$\times$ & 0.62$\times$ & 0.82$\times$ & 1.00$\times$ & 0.82$\times$ & 1.00$\times$ & 1.00$\times$ & 1.00$\times$ & 0.99$\times$ & 1.00$\times$ & 1.01$\times$ \\
\rowcolor{colB} \textbf{SSSP} & \textcolor{mygreen}{0.56$\times$} & \textcolor{mygreen}{0.54$\times$} & 1.70$\times$ & 0.55$\times$ & 0.94$\times$ & 2.50$\times$ & 0.00$\times$ & 1.00$\times$ & 152.74$\times$ & 1.76$\times$ & 2.00$\times$ & 1.17$\times$ & 0.53$\times$ & 0.63$\times$ & 1.25$\times$ & 0.00$\times$ & 1.00$\times$ & 137.67$\times$ & 1.82$\times$ & 2.01$\times$ \\
\rowcolor{colA} \textbf{SSSP-I} & \textcolor{mygreen}{0.56$\times$} & \textcolor{mygreen}{0.54$\times$} & 1.70$\times$ & 0.55$\times$ & 0.94$\times$ & 2.50$\times$ & 0.00$\times$ & 1.00$\times$ & 140.56$\times$ & 1.76$\times$ & 1.99$\times$ & 1.28$\times$ & 0.54$\times$ & 0.69$\times$ & 1.50$\times$ & 0.00$\times$ & 1.00$\times$ & 289.83$\times$ & 1.81$\times$ & 1.97$\times$ \\
\midrule
\rowcolor{colB} \textbf{CSV-V} & \textcolor{mygreen}{0.87$\times$} & \textcolor{mygreen}{0.92$\times$} & 0.69$\times$ & 1.06$\times$ & 0.73$\times$ & 0.91$\times$ & 0.66$\times$ & 0.67$\times$ & 1.00$\times$ & 0.08$\times$ & 0.74$\times$ & 0.76$\times$ & 0.95$\times$ & 0.73$\times$ & 0.91$\times$ & 0.66$\times$ & 0.67$\times$ & 1.00$\times$ & 0.79$\times$ & 0.65$\times$ \\
\rowcolor{colA} \textbf{PR} & \textcolor{myred}{1.40$\times$} & \textcolor{myred}{1.24$\times$} & 1.07$\times$ & 1.40$\times$ & 1.50$\times$ & 2.10$\times$ & 1.05$\times$ & 2.01$\times$ & 1.00$\times$ & 1.11$\times$ & 1.05$\times$ & 0.78$\times$ & 1.24$\times$ & 0.96$\times$ & 1.02$\times$ & 1.04$\times$ & 1.49$\times$ & 0.99$\times$ & 1.10$\times$ & 1.04$\times$ \\
\rowcolor{colB} \textbf{SL-I} & \textcolor{mygreen}{0.75$\times$} & \textcolor{mygreen}{0.73$\times$} & 3.38$\times$ & 0.74$\times$ & 2.51$\times$ & 3.37$\times$ & 3.80$\times$ & 2.46$\times$ & 1.00$\times$ & 0.72$\times$ & 0.60$\times$ & 3.18$\times$ & 0.73$\times$ & 2.31$\times$ & 2.86$\times$ & 3.71$\times$ & 2.14$\times$ & 1.00$\times$ & 0.71$\times$ & 0.62$\times$ \\
\rowcolor{colA} \textbf{SL-R} & \textcolor{mygreen}{0.69$\times$} & \textcolor{mygreen}{0.69$\times$} & 3.70$\times$ & 0.66$\times$ & 2.45$\times$ & 3.51$\times$ & 4.58$\times$ & 2.06$\times$ & 0.98$\times$ & 0.79$\times$ & 0.53$\times$ & 3.34$\times$ & 0.66$\times$ & 2.21$\times$ & 3.03$\times$ & 4.42$\times$ & 1.92$\times$ & 0.96$\times$ & 0.79$\times$ & 0.54$\times$ \\
\rowcolor{colB} \textbf{SL-S} & \textcolor{mygreen}{0.68$\times$} & \textcolor{mygreen}{0.68$\times$} & 2.72$\times$ & 0.68$\times$ & 1.86$\times$ & 2.71$\times$ & 6.92$\times$ & 1.78$\times$ & 0.99$\times$ & 0.64$\times$ & 0.56$\times$ & 2.54$\times$ & 0.68$\times$ & 1.73$\times$ & 2.35$\times$ & 6.92$\times$ & 1.64$\times$ & 0.98$\times$ & 0.63$\times$ & 0.56$\times$ \\
\rowcolor{colA} \textbf{u8next} & \textcolor{mygreen}{0.67$\times$} & \textcolor{mygreen}{0.76$\times$} & 1.57$\times$ & 0.70$\times$ & 1.09$\times$ & 1.12$\times$ & 1.00$\times$ & 1.00$\times$ & 0.91$\times$ & 0.79$\times$ & 0.04$\times$ & 1.43$\times$ & 0.77$\times$ & 1.09$\times$ & 1.12$\times$ & 1.00$\times$ & 1.00$\times$ & 0.86$\times$ & 0.79$\times$ & 0.04$\times$ \\
\bottomrule
\end{tabular}%
}
\caption{Rust/C performance ratios (AMD). Benchmark
abbreviations follow~\autoref{tab:bench_perf_summary}.
Time values $\leq 1.00\times$ are in \textcolor{mygreen}{green}
and values $\geq 1.10\times$ are in \textcolor{myred}{red}.}
\label{tab:ratios_amd}
\end{table*}
We now discuss the benchmark overheads using the microarchitectural
characteristics and dynamic instruction traces.
For all but 4 benchmarks, at least one Rust implementation had 
comparable (within 10\%) performance or \emph{outperformed} C. 
Benchmarks involving few bounds checks, such as graph traversals, are natural candidates for low overhead. Indeed, BFS incurs only 10\% overhead on AMD (23\% on Intel), and DFS only 5\% on both architectures.
However, we observe that surprisingly,
even for some benchmarks that
contain a high number of bounds checks in tight loops (SSSP, DFS, and Quicksort), the 
Rust overhead is less than 10\%.
Below, we discuss the microarchitectural
components that hide Rust's overheads
(\autoref{tab:ratios_amd} and \autoref{tab:ratios_intel}).

\subsubsection{Impact of Branch Predictors}
\label{sec:hiding_bp}
Indexing into a \texttt{vec} or a \texttt{slice} invokes a 
bounds-check consisting of three operations---i) loading
the size of the \texttt{vec} or \texttt{slice} object, ii) comparing the index
with the size, and finally iii) depending on the comparison
result retrieving the element at the index or panicking.
Thus, if the vector's size is not promoted to a register
by the Rust compiler and results in a long latency memory 
load operation to resolve the branch condition, it can stall the CPU pipeline.
Therefore, any benchmark 
involving frequent bounds checks, 
such as random index-based sorting benchmarks,
could be expected to incur a high overhead. 
Yet, across the 18 benchmarks and variants, branch predictor misses \emph{decreased} or stayed within 10\% of the baseline C version in most cases—14 of 18 on AMD and 13 of 18 on Intel, showing
that the hardware
branch predictor can typically accurately predict the branch targets of 
the bounds checks branches. 

\para{Cases where BPs Hide the Costs.} We 
observe that for the DFS and Quicksort benchmarks, despite incurring a roughly 60\% increase in conditional branches on both architectures, the number of branch misses remains within 3\% of the C implementation or less. Both of these cases result in at most a 7\% increase in runtime overhead overall. More impressively, the Skip List operations show a drastic increase in branch instructions, ranging from 2.71$\times$ for the Search operation to 3.51$\times$ for the Remove operation, but show
an equal or better branch miss count to the C baseline.
Combined with improved cache performance, discussed in \autoref{sec:hiding_cache}, this results in an average 30\% and 6\% \emph{runtime improvement} on AMD and Intel, respectively.

\para{Iterators Elide Bounds Checks.}
Unlike index-based accesses, using iterators
to traverse a \texttt{vec} or \texttt{slice} variable,
does not involve bounds checks at all.
Our results show that the use of iterators can \emph{generally}
aid in reducing bounds checking overhead. For example, the iterator-based Matrix Multiplication benchmark successfully removed
nearly all bounds checks and saw only a 1\% increase in the dynamic branch count over C
and equal branch misses, resulting in a 5\% \emph{runtime improvement} for AMD and a small 1\% decrease for Intel. 
A similar pattern holds for the iterator-based, top-down recursive Merge Sort implementation.
While it incurs a modest increase in dynamic memory and branch instructions, it 
almost completely eliminates branch mispredictions to just 0.09$\times$ and 0.05$\times$ for AMD and Intel, respectively, compared to the C baseline. This is a striking difference compared to the $\sim$18$\times$ increase in branch misses seen
in the index-based Top-Down, Recursive variant.
Combined with competitive cache performance, the iterator implementation ultimately \emph{outperforms} C with runtime overheads of 0.85$\times$ for AMD and 0.66$\times$ for Intel.
We note that compared to the baseline C version, the iterator-based Merge Sort variant \emph{increases} the number of conditional checks 
to be performed. These checks stem from iterator semantics ensuring that the \texttt{left}, \texttt{right}, and output array iterators are 
still viable in the \texttt{Merge} function’s hot loop. In contrast, the C code only checks if the \texttt{left} and \texttt{right} array indices are still valid, 
resulting in 2 instead of 3 checks.

\para{Cases Where BPs Cannot Hide the Costs.}
We observed that for benchmarks such 
as Merge Sort and SSSP, the Rust variants
exhibit a \emph{significant increase}
in the number of executed branch instructions,
resulting in an increased pressure on the
branch predictors. 
Thus, the branch mispredictions are naturally higher for Rust than C.
This is particularly evident in the Merge Sort benchmark. Both the top-down recursive (TDR) and bottom-up iterative (BUI) implementations
exhibit a substantial increase in dynamic branch instructions for Rust over C: 2.13$\times$ for TDR and 2.69$\times$ for BUI on both architectures. This increase 
corresponds to a rise in branch mispredictions on both architectures, approximately 18$\times$ for the top-down recursive 
implementation and 20$\times$ for the bottom-up iterative implementation. 
We discuss
the compiler behavior responsible for this
in \autoref{sec:compilereffects}.

\subsubsection{Impact of Data Caches}    
\label{sec:hiding_cache}
Across both architectures, the Rust implementations of Depth-First Search, SSSP, and Skip List ranges from
21–56\% fewer L1D-cache misses compared to the C baseline.
On the Intel machine specifically, Quicksort also follows this trend, while Binary Search, LibCsv, and UTF-8 Decode demonstrate even more significant reductions exceeding 50\% (up to 97\% for UTF-8 Decode).
Conversely, on the AMD machine, Breadth-First Search, Quicksort, and Selection Sort also display L1D-cache behavior comparable to their C counterparts.
In most cases, these cache-efficiency gains occur despite the Rust implementations issuing more memory instructions overall.
This shows that even if the 
Rust code compiles to more memory operations, potentially due to 
memory-safety checks, the L1D-cache can
capture these accesses, thus limiting their overhead.

To illustrate, the Rust translation of the Skip List search operation issues 1.78$\times$ more memory load operations and 6.92$\times$ more store operations than the C baseline.
However, the L1D-cache misses
\emph{decreases} by 36\%, and L2 misses
by 44\%, resulting in an effective 32\% \emph{decrease}
in the runtime compared to the baseline C version on the AMD machine. A more modest but similar pattern is found on the Intel machine. Similarly,
the Rust version of the Binary Search benchmark, which has only a 11\% increase
in the dynamic memory store instructions, has a 73\% decrease in the
L1D-cache miss rate on the AMD machine.
The Single-Source Shortest Path benchmark was
the only benchmark to exhibit a drastic decrease in dynamic
memory instructions accompanied by a high \emph{increase}
in cache misses. For SSSP, the L1D-cache misses increase by 76\% over the baseline C, but still result in a runtime improvement of 44\% on AMD.
The behaviors of both the Binary Search and SSSP benchmarks are further influenced by diverging compiler optimization effects, which we explore further in \autoref{sec:compilereffects}.

\begin{table}[t]
\centering
\footnotesize
\setlength{\tabcolsep}{3pt}
\resizebox{\linewidth}{!}{%
\begin{tabular}{@{}rrrrr@{\hspace{1em}}rrrrr@{}}
\toprule
\multicolumn{5}{c}{\textbf{Prefetches}} &
\multicolumn{5}{c}{\textbf{L1D Cache Misses}} \\
\cmidrule(lr){1-5} \cmidrule(lr){6-10}
\textbf{C} &
\textbf{Rust (vec)} &
\textbf{\rmm{} (vec)} &
\textbf{Rust (raw)} &
\textbf{\rmm{} (raw)} &
\textbf{C} &
\textbf{Rust (vec)} &
\textbf{\rmm{} (vec)} &
\textbf{Rust (raw)} &
\textbf{\rmm{} (raw)} \\
\midrule
962.35 & 225.16 & 225.61 & 225.20 & 225.23 &
1802.42 & 478.99 & 479.06 & 479.00 & 479.24 \\
\bottomrule
\end{tabular}%
}
\caption{Hardware prefetches and L1D-cache misses
(in millions) for C, Rust, \rmm{}, and raw-pointer
variants of Binary Search on AMD.}
\label{tab:binary-search-prefetch}
\end{table}

\para{Higher Number of Prefetches in C.}
To analyze the differing cache behavior between C and Rust, we conducted
a series of experiments with the Binary Search benchmark. We first
replaced the Rust vector with a raw pointer, which would ensure it would
perform no bounds checks. As shown in the results in 
\autoref{tab:binary-search-prefetch}, both the vector-based and raw-pointer based
implementations have significantly lower L1D-cache miss rates.
The only difference we could find between the C and Rust versions was
that the C code was making $\sim$5$\times$ as many prefetch requests as the Rust
baseline.
By disassembling the binary, we confirmed that these were hardware 
prefetcher requests, and because we could not control the BIOS settings
on the CloudLab machines, we could not turn it off to further isolate the
behavior. Furthermore, the
hardware performance counter
records only the L1D-cache 
misses resulting from \emph{demand accesses} and not prefetches. 
Therefore, we hypothesize that the higher
number of prefetches in C code evicts \emph{hot} data in the cache,
resulting in a higher L1D-cache miss rate.

\subsection{Compiler Optimization Impact}
\label{sec:compilereffects}

As previously noted by Popescu et al.~\cite{popescu2021safer}, the runtime
safety checks can alter compiler optimizations in unpredictable ways. 
We discuss some of the interesting compiler optimization behaviors
we observed during our analysis.

\subsubsection{Promotion of Bounds Value to Registers}
We observed that for the Index-based Quicksort and the Selection Sort
algorithms, despite performing additional memory-safety checks, as evidenced
by the increase in dynamic branch instructions, there is little increase in the 
memory load instructions, with a 2\% increase for Quicksort and no change for Selection Sort.
This indicates that the compiler
could successfully promote the 
\texttt{vector} size to a register,
reducing any latency related to memory accesses. However, we
did not observe this behavior in any of the other benchmarks, indicating
that either some of the bound values could not be promoted to registers,
or the extra register pressure from the bounds values
caused other frequently-accessed variables to be spilled to the stack.

\begin{table}[t]
\centering
\footnotesize
\setlength{\tabcolsep}{3pt}

\begin{minipage}[t]{0.32\linewidth}
\vspace{0pt}
\centering
\begin{tabular}{@{}lrrr@{}}
\toprule
\textbf{App.} & \textbf{C} & \textbf{Rust} & \textbf{\rmm{}} \\
\midrule
BFS    & 9     & 60.0k  & 60.0k \\
BSrch  & 1     & 892.9M & 892.9M \\
DFS    & 0     & 0      & 0 \\
MM     & 0     & 0      & 0 \\
MM-I   & 4.1k  & 4.0M   & 4.0M \\
MS-TDR & 2.53B & 1.02B  & 2.94B \\
MS-BUI & 2.74B & 1.00B  & 2.84B \\
\bottomrule
\end{tabular}
\end{minipage}%
\hfill
\begin{minipage}[t]{0.32\linewidth}
\vspace{0pt}
\centering
\begin{tabular}{@{}lrrr@{}}
\toprule
\textbf{App.} & \textbf{C} & \textbf{Rust} & \textbf{\rmm{}} \\
\midrule
MS-I   & 2.53B & 3.17B  & 3.17B \\
QS     & 0     & 0      & 38.6M \\
QS-I   & 0     & 125.0M & 125.0M \\
SelS   & 0     & 0      & 0 \\
SSSP   & 1.35B & 5      & 5 \\
SSSP-I & 1.35B & 4      & 4 \\
\bottomrule
\end{tabular}
\end{minipage}%
\hfill
\begin{minipage}[t]{0.32\linewidth}
\vspace{0pt}
\centering
\begin{tabular}{@{}lrrr@{}}
\toprule
\textbf{App.} & \textbf{C} & \textbf{Rust} & \textbf{\rmm{}} \\
\midrule
CSV-V  & 12.4M & 20.1M  & 20.1M \\
PR     & 4     & 103    & 0 \\
SL-I   & 15.0M & 37.5M  & 37.5M \\
SL-R   & 1     & 497    & 497 \\
SL-S   & 0     & 5      & 5 \\
u8next & 0     & 0      & 0 \\
\bottomrule
\end{tabular}
\end{minipage}

\caption{Dynamic CMOV instruction counts for C, Rust,
and \rmm{}. Counts are identical across AMD and Intel. }
\label{tab:cmov}
\end{table}

\subsubsection{Bounds Checks Preventing Loop Unrolling.}
\label{sec:unroll-prevent}Popescu et.al~\cite{popescu2021safer}
hypothesized that Rust's safety checks can inhibit
other compiler optimizations. We found an evidence of that behavior for the Selection Sort benchmark.
For both the original C and the \rmm{} versions, the LLVM compiler can successfully unroll the core
loop 4$\times$, thus reducing the branch
instructions. However, the \sysname debugger
identified that this loop unrolling optimization
was not successful in the Rust version.
Using the LLVM compilers \emph{CRemark}~\cite{llvm-remarks}
feature, the \sysname debugger identified
a provably redundant bounds check in the inner loop that searches for
the minimum element. 
The loop compares
\texttt{arr[j] < arr[min\_idx]} and updates
\texttt{min\_idx} to \texttt{j} when the comparison holds.
Since \texttt{min\_idx} is always either the outer loop
index \texttt{i} or some \texttt{j < len}, the bounds check
can never fail. A simple check between possible values of \texttt{min\_idx} (the ranges of \texttt{i} and \texttt{j}) and the length off \texttt{arr[]} would prove this check redundant. However, the Rust compiler does not
eliminate it, introducing multiple panic exits within
the loop. These panic exits ultimately
prevent the $4\times$ loop unrolling
optimization observed in both C and \rmm{}.
Similarly, Rust's bounds checks also 
prevents the \texttt{CMOV} optimization which we 
discuss in~\autoref{sec:cmov}.

\subsubsection{Converting Branch + Load Patterns to Conditional Moves}
\label{sec:cmov}
One possible compiler optimization is to turn conditional branch instructions followed
by a memory operation---for
example the \texttt{CMP rdi, rdx; JNE .LBB19\_1; MOV rax, ptr} sequence shown in \autoref{fig:safetycheck}, with 
a conditional move instruction, such as \texttt{CMP rdi, rdx; CMOV rax, ptr;}.
The \texttt{CMOV}
instruction performs a memory move operation if the previous condition evaluates to a certain
value.
This reduces the number of executed instructions
and essentially converts a control dependency, in the form of the branch instruction, 
into a data dependency. 
Our results show this optimization has a mixed 
impact on performance.
\autoref{tab:cmov} shows how often this 
transformation occurs across the 
benchmarks. 

\para{Missed CMOV Optimization in Rust.}
For the index-based variants of the Merge Sort benchmark,
the C compiler applies the 
conditional load patterns to  \texttt{CMOV} promotion, which reduces the overall number of branches, and also the branch
predictor 
misses~(\autoref{tab:ratios_intel},~\autoref{tab:ratios_amd}).
However, the Rust compiler failed
to apply this optimization to
the Rust code, resulting in a significantly
higher number of mispredictions ($\sim$18$\times$ TDR, $\sim$20.5$\times$ BUI).
In Merge Sort (TDR), \sysname identifies that the \texttt{CMOV}-conversion is being prevented by Rust's bounds checks.
In the Rust code, the bounds check on \texttt{arr[k]} is performed separately in
each branch of the main merge loop's comparison
(\texttt{left\_arr[i] <= right\_arr[j]}). Each check has a call to \texttt{panic} which provides different \texttt{Location} information including line number and file. 
If a check fails, execution exits through its corresponding panic path.
As a result, these two checks cannot be merged into one and hoisted.
For Merge Sort (BUI), \sysname's Optimization Checker
infers that the compiler's CFG simplification
pass is unable to hoist the common test or merge the two stores into a select because of this. 
\autoref{tab:cmov} shows 
that in the case of both indexed (non-iterator) implementations of Merge Sort, compiling with \rmm{} enabled the same CMOV optimization that C used and made a significant improvement in performance, even surpassing C's performance for the bottom up (iterative) implementation.

\para{CMOV Misoptimization in Rust.}
The \texttt{CMOV} optimization does not always lead to improved performance.
As discussed in \autoref{sec:hiding_bp}, the hardware branch predictor
is very efficient for predictable branch checks, so this conversion likely 
is inefficient and is, in fact, a mis-optimization for benchmarks such as Binary Search. 
Binary Search takes 1.26$\times$ C's time in Rust, and removing runtime checks has no effect (1.25$\times$ in \rmm{}). 
Using the LLVM compiler's CRemarks~\cite{llvm-remarks} framework, the
\sysname debugger's Optimizer Checker traced this overhead 
to the LLVM compiler's \emph{Instruction Selection} pass. 
The LLVM compiler's x86-64 backend 
lowers the comparisons of \texttt{mid} with 
\texttt{lo} and \texttt{hi} with a branch in C but with \texttt{CMOV}s in Rust (\autoref{tab:cmov}). 
The \texttt{CMOV}s make the selection of
the next midpoint, and therefore the next load 
operation, data-dependent on the current load, 
requiring so consecutive loads to be serialized. 
\sysname's $\mu$arch Checker verified this with hardware counters. 
Rust has 0.06$\times$ C's branch misses and 0.27$\times$ its cache misses, yet needs 1.27$\times$ the cycles (\autoref{tab:ratios_amd}). 

We disabled CMOV
conversion with the flag \texttt{-x86-cmov-converter-force-all=true} for SSSP in C and Binary Search
in Rust. We found that without CMOV, SSSP was able to reduce its runtime by 57\%, outperforming the
Rust implementation. Removing CMOVs for Binary 
Search in Rust reduced its overhead to only 7\% 
overhead over C.

\para{CMOV Misoptimization in C.}
The CMOV misoptimization, however, is not exclusive to Rust. In the SSSP benchmark, the C implementation is similarly misoptimized using \texttt{CMOV}. When updating the \texttt{distances} array, which stores the shortest known distance from the source vertex to every other vertex,
the C compiler computes the minimum using a \texttt{CMOV} instruction, whereas the Rust compiler emits a conditional branch followed by a memory store.
While this difference leads to increased branches and therefore more branch mispredictions compared to C, it reduces the number of store instructions from over 1B in C to just over 200K in Rust, ultimately resulting in a 32\% to 44\% \emph{increase} in performance for Rust.

\subsubsection{Aliasing Impact}
Rust's ownership model also affects how aggressively LLVM can optimize
memory accesses. Compilers, such as LLVM~\cite{llvm} and GCC~\cite{gcc}, use \emph{alias analysis
(AA)}~\cite{steens, andersen} to determine whether two pointers can refer to the same memory. When the
compiler can prove they do not (NoAlias), it can better optimize the memory accesses by reordering or
vectorizing them aggressively. LLVM conveys aliasing information through a combination of attributes,
annotations, and metadata, which we collectively refer to as \emph{alias information}. Rust's borrow
checker guarantees that no mutable reference coexists with any other reference to the same memory, and
the compiler encodes this by emitting \texttt{noalias} alias information for \texttt{\&mut T} and
\texttt{\&T} pointer parameters in the LLVM IR, which becomes scoped \texttt{noalias} metadata on the
individual loads and stores when a function is inlined. C pointers carry no such alias information
unless explicitly annotated with the 
\texttt{restrict} keyword. Instead, C relies on its strict-aliasing rules,
which Clang encodes as type-based alias analysis (TBAA) metadata on each load and store, letting LLVM
separate accesses of incompatible types~\cite{llvm-tbaa, diwan1998tbaa}, while 
\texttt{rustc} emits no TBAA.
One might expect Rust's ownership-based
alias information to result in more NoAlias relationships
as hypothesized by Villani et
al.~\cite{villani2025treeborrows}. However, we observe that this behavior is dependent on how the code
is implemented and how memory is allocated.

To characterize the impact of Rust's type system on aliasing information, we executed LLVM's Alias
Analysis~\cite{llvm-aa-eval} pass (\texttt{aa-eval}) on the O3-optimized IR of the hot functions of
PageRank and Matrix Multiplication (\autoref{tab:alias}). We report the pass's load/store queries
(\texttt{--evaluate-aa-metadata}), which take each access's alias information into account, including
TBAA and scoped \texttt{noalias} metadata. 
The \texttt{aa-eval} pass's default pointer-pair queries omit this
metadata and therefore understate NoAlias for both languages.

\para{Slice Type Improves Rust's AA precision.}
In PageRank, 24.4\% of Rust's queries are NoAlias, compared to 14.5\% for C. This gap arises because
the C function takes five raw pointers, of which three are of \texttt{float*} type. TBAA separates the
\texttt{float} accesses from the integer accesses, but it cannot separate pointers of the same type, so
accesses through the three \texttt{float*} parameters remain MayAlias. The Rust version's three
immutable slice parameters all carry \texttt{noalias} alias information due to its stronger type
system, which resolves exactly these pairs.

\para{Nested Vectors Limit Rust's AA precision.}
For Matrix Multiplication, \sysname's Phase~2 identified a disconnect between the guarantees provided by Rust's borrowing rules and what can be communicated to LLVM through \texttt{noalias} annotations.
For nested \texttt{Vec} parameters (\texttt{Vec<Vec<i32>{}>}), \texttt{noalias} applies only to the outer \texttt{Vec} references and cannot convey non-aliasing information about the inner \texttt{Vec}s.
As a result, LLVM must conservatively assume that writes through an inner \texttt{Vec} could alias its header.
Matrix Multiplication shows the effect of this on alias analysis. 
In the original \texttt{i-j-k} implementation, none of Rust's queries are NoAlias, while C resolves 40.0\% of them. Each store \texttt{result[i][j]} writes through a row
pointer loaded from the outer \texttt{Vec<Vec<i32>{}>}. Rust's \texttt{noalias} covers only the
references passed to the function, not the rows reached through them, so LLVM cannot rule out that
the store overwrites a row header, and it reloads the header and keeps the accumulator in memory. In
C, TBAA marks the stores as \texttt{int} accesses and the row pointers as \texttt{int*} accesses,
which lets LLVM separate them. We modified the program so that the inner computation is a separate
kernel function that takes a shared slice of the row and a mutable reference to the output element.
The output address then arrives as a \texttt{noalias} parameter instead of being loaded from memory
inside the loop, and Rust's NoAlias share rises to 18.2\%. \sysname flagged similar situations in the BFS and Pagerank benchmarks.

\begin{table}[t]
\centering

\begin{minipage}[t]{0.49\linewidth}
  \centering
  \setlength{\tabcolsep}{2pt}
  \resizebox{\linewidth}{!}{%
  \begin{tabular}{@{}lrrrrrr@{}}
  \toprule
  & \multicolumn{2}{c}{\textbf{PageRank}}
  & \multicolumn{2}{c}{\textbf{Matrix Mult.}}
  & \multicolumn{2}{c}{\makecell{\textbf{Matrix Mult.}\\(Mod.)}} \\
  \cmidrule(lr){2-3}
  \cmidrule(lr){4-5}
  \cmidrule(lr){6-7}
  & \textbf{C} & \textbf{Rust}
  & \textbf{C} & \textbf{Rust}
  & \textbf{C} & \textbf{Rust} \\
  \midrule
  \textbf{Total Queries} & 297 & 180 & 50 & 42 & 35 & 22 \\
  \textbf{NoAlias}   & 14.5\% & 24.4\% & 40.0\% & 0.0\%  & 31.4\% & 18.2\% \\
  \textbf{MayAlias}  & 84.2\% & 73.3\% & 48.0\% & 97.6\% & 51.4\% & 77.3\% \\
  \textbf{MustAlias} & 1.3\%  & 2.2\%  & 12.0\% & 2.4\%  & 17.1\% & 4.5\% \\
  \bottomrule
  \end{tabular}%
  }
  \caption{Alias analysis results (\texttt{aa-eval}
  load/store queries with alias metadata on
  O3-optimized hot functions). Matrix Mult.\
  (Mod.) sums the kernel and its caller.}
  \label{tab:alias}
\end{minipage}%
\hfill
\begin{minipage}[t]{0.49\linewidth}
  \centering
  \setlength{\tabcolsep}{2pt}
  \resizebox{\linewidth}{!}{%
  \begin{tabular}{@{}lrrrrrr@{}}
  \toprule
  & \multicolumn{3}{c}{\textbf{Matrix Mult.}}
  & \multicolumn{3}{c}{\makecell{\textbf{Matrix Mult.}\\(Mod.)}} \\
  \cmidrule(lr){2-4}
  \cmidrule(lr){5-7}
  & \textbf{C} & \textbf{Rust} & \textbf{\rmm{}}
  & \textbf{C} & \textbf{Rust} & \textbf{\rmm{}} \\
  \midrule
  \textbf{Runtime (s)}
    & 13.38 & 25.54 & 23.06
    & 13.45 & 19.65 & 14.14 \\
  \textbf{Overhead}
    & -- & 90.8\% & 72.3\%
    & -- & 46.0\% & 5.1\% \\
  \bottomrule
  \end{tabular}%
  }
  \caption{Matrix multiplication runtimes at
  $N=2000$.
  Overhead is relative to C in each group.}
  \label{tab:matrix-runtime}
\end{minipage}

\end{table}

\para{Aliasing Performance Impact.}
Since alias information can affect code generation, it has the potential to affect performance. The matrix multiplication example shows this. The alias information gained by keeping the kernel as a separate function with the output address known
not to alias the row headers allows LLVM promote the accumulator to a register, closer matching C's code generation.
\autoref{tab:matrix-runtime} shows that Rust's overhead falls from 90.8\% to 46.0\%, and that
\rmm{}, which also drops the bounds checks, comes within 5.1\% of C.

\subsection{\texttt{Rc<RefCell<>{}>} Overhead} We compare Rust's Skiplist implementation to a version that
uses the \texttt{unsafe} \texttt{NonNull} raw pointer abstraction instead of \texttt{Rc<RefCell<>{}>}
in its primary data structure. Running these through a modified version of \sysname that allows for
Rust-to-Rust comparison shows that \texttt{Rc<RefCell<>{}>} costs 11--17\% per phase
(Table~\ref{tab:skiplist-runtime}). The candidate executes 1.7--2.0$\times$ the instructions of the
raw-pointer version for the same number of comparisons. Most of the extra work is inherent to the
abstraction. Each node visited during a traversal takes a reference-count increment and decrement and
two \texttt{RefCell} borrow-flag checks, and all of these write to the node being visited. Because the
traversal is a latency-bound pointer chase, much of this work overlaps with existing memory stalls,
and IPC rises from about 0.3 to 0.5. The rest of the overhead comes from layout. The
\texttt{Rc<RefCell<>{}>} node is 64 bytes instead of 40, so its hot fields cross a cache-line boundary
more often, and L2 misses in the remove phase rise by 21\%. \sysname also found two missed
optimizations. Handles that could stay in registers are stored to and reloaded from the stack on every
step, because their addresses escape to an out-of-line \texttt{drop\_slow} whose declaration lacks
\texttt{captures(none)}. Some links are also reloaded because alias information is lost through
\texttt{Rc} and \texttt{Cell}. Together these account for an estimated 15--20\% of the extra
instructions, but only a few percent of the runtime. Both Rust versions are faster
than the C implementation in every phase.

\section{Compiler Pass for Bounds Check Hoisting}

To illustrate the usefulness of our workload
characterization we leverage the gained insights
into a compiler pass.
Our manual analysis reveals that many bounds checks in our benchmarks are unnecessary, 
but their presence can prevent compiler optimizations that substantially improve performance.
We therefore implement an LLVM pass that removes these checks while preserving the precise semantics of Rust bounds-check panics. 
That is, an out-of-bounds access must still panic at the exact location and time at which it occurs.
We argue that such a pass allows programmers to use Rust's natural indexing operations without worrying about the performance overhead that bounds checks can introduce.
The pass derives sound, loop-invariant conditions under which checks are guaranteed to succeed and uses them to version the loop, 
executing a copy without checks when the conditions hold and retaining the original loop as a fallback otherwise. \autoref{sec:related-works} discusses how our pass improves over prior related work.

\subsection{Approach.}
The pass treats a conditional branch in a loop $L$ as a bounds check if its failing result leaves $L$ and reaches a call that does not return, such as a \texttt{panic\_bounds\_check}.
The pass then expresses each check $c$ as a comparison $x \bowtie B$ that is
true when the check passes, where $\bowtie$ is an integer comparison
($<$, $\le$, $>$, $\ge$ or $\ne$, signed or unsigned) and $B$ is loop invariant.
It derives a loop-invariant predicate $\varphi_c$ that implies $c$ on every iteration of $L$.
It then \emph{versions} $L$, where the preheader evaluates $\Phi = \bigwedge_c \varphi_c$ (a conjunction of each bounds check condition) once and
enters a copy of $L$ with every check removed if $\Phi$ holds, or the unchanged original loop otherwise.
To derive $\varphi_c$, the pass tries four rules and uses the first whose operands it can compute in the preheader.

\para{Rule 1: Induction Variables.} Suppose $x$
is an induction variable that starts at $s$ and 
changes by the same step $d$ on every iteration,
like a loop counter. Then $x$ only moves in one direction, so if the check
passes for $x$'s first and last values, it passes for every value in between.
If the loop runs at most $N$ iterations, $x$ never goes past
$s + d\cdot(N-1)$, so
$\varphi_c \equiv (s \bowtie B) \wedge (s + d\cdot(N-1) \bowtie B)$. 
This holds only if $x$ does not overflow or underflow,
since a value that wraps around no longer moves in one direction. 
The pass therefore computes $s + d\cdot(N-1)$ in a wider integer type, where it is exact 
and adds to $\varphi_c$ the condition that the result fits in $x$'s type.
To find $N$, the pass considers every exit that the loop tests on every iteration, other than the checks
themselves, and takes the smallest iteration count among them, since the loop
must stop at whichever exit comes first. 
Exits only taken on some paths, such as within a nested if, are not considered.

\para{Rule 2: Leveraging Existing Guard Conditions.} The pass then considers the case that the code already limits $x$ before the check
  runs. For example, if every path to the check first passes a test that guarantees
  $x < M$, where $M$ does not change in the loop, then the check $x < B$
  passes whenever $M \le B$, so $\varphi_c \equiv M \le B$.
 
\para{Rule 3: Conditionally Assigned Variables.} The pass
considers the case that $x$ is chosen conditionally. 
On each iteration some variable $x$ either keeps its old value or takes a new value. 
For example, in Selection Sort's
  \texttt{if arr[j] < arr[min\_idx] \{ min\_idx = j; \}}. In this example, $x$ always holds
  one of its candidate values. \texttt{min\_idx} starts as \texttt{i} and only
  ever becomes some \texttt{j}. If each candidate either does not change in the
  loop or is covered by rule~1, then the check on $x$ passes whenever it passes
  for every candidate, so $\varphi_c$ is the conjunction of the candidates'
  predicates. For \texttt{arr[min\_idx]}, that is \texttt{i < len} together
  with the predicate for \texttt{arr[j]}.
  
\para{Rule 4: Counter Sum Bounds Analysis.} Finally,
the pass considers the case where some variable $x$ advances only
  together with other counters. Suppose a loop has counters $i$, $j$ and $k$,
  where $i$ advances on iterations where some condition holds, $j$ advances on
  the remaining iterations, and $k$ advances on every iteration, as in the hot
  loop of merge sort's merge function. Then $k$ has always grown by exactly as
  much as $i$ and $j$ together. If rule~2 bounds $i$ by $M_i$ and $j$ by
  $M_j$, for example through the loop condition, then $k$ never exceeds
  $k_0 + (M_i - i_0) + (M_j - j_0)$, where $k_0$, $i_0$ and $j_0$ are the
  starting values, and
  $\varphi_c \equiv k_0 + (M_i - i_0) + (M_j - j_0) < B$. In general, the
  pass requires that on every path through the loop body $x$ grows by no more
  than the other counters grow together, and that none of them decreases.
  Rule~1 cannot handle $k$ since it needs the number of iterations, which LLVM
  cannot compute when the loop ends on counters that advance irregularly.

\subsection{Results.} \autoref{tab:sli-runtime}
shows the results. With the checks gone, LLVM successfully converts the branches in the merge loop of both merge sorts into \texttt{CMOV} instructions as C does, discussed in \autoref{sec:cmov}.
Merge Sort (BUI) then runs slightly faster than C and on-par with \rmm{}
(\autoref{tab:ratios_amd}). Merge Sort (TDR) retains 17\% overhead, close to \rmm{}'s 20\% seen in \autoref{tab:ratios_amd}. For
Selection Sort, the loop without checks is
unrolled 4$\times$ like C's, discussed in \autoref{sec:unroll-prevent}, and the remaining overhead disappears. The pass
has no effect on Matrix Multiplication since the bound of \texttt{matrix\_b[k][j]} is the length of row
\texttt{k}, which changes on every iteration of the hot loop and cannot be proven in the preheader.

\begin{table}[t]
\centering
\footnotesize

\begin{minipage}[t]{0.49\linewidth}
  \vspace{0pt}
  \centering
  \setlength{\tabcolsep}{3pt}
  \resizebox{\linewidth}{!}{%
  \begin{tabular}{@{}lrrr@{}}
  \toprule
  & \makecell{\textbf{Rust}\\(\texttt{NonNull})}
  & \makecell{\textbf{Rust}\\(\texttt{Rc<RefCell<>{}>})}
  & \textbf{Overhead} \\
  \midrule
  \textbf{Insert} & 15.80 & 18.43 & 16.6\% \\
  \textbf{Remove} & 8.07  & 9.35  & 15.9\% \\
  \textbf{Search} & 4.66  & 5.19  & 11.2\% \\
  \bottomrule
  \end{tabular}%
  }
  \caption{Skiplist runtimes in seconds per phase. Overhead is
  \texttt{Rc<RefCell<>{}>} relative to
  \texttt{NonNull}.}
  \label{tab:skiplist-runtime}
\end{minipage}%
\hfill
\begin{minipage}[t]{0.49\linewidth}
  \vspace{0pt}
  \centering
  \setlength{\tabcolsep}{3pt}
  \resizebox{\linewidth}{!}{%
  \begin{tabular}{@{}lrrr@{}}
  \toprule
  & \textbf{C} & \textbf{Rust} & \textbf{Rust + pass} \\
  \midrule
  \textbf{Merge Sort (BUI)}
    & 10.53 & 15.02 (1.43$\times$) & 9.55 (0.91$\times$) \\
  \textbf{Merge Sort (TDR)}
    & 13.10 & 21.68 (1.65$\times$) & 15.39 (1.17$\times$) \\
  \textbf{Selection Sort}
    & 6.81 & 20.09 (2.95$\times$) & 6.80 (1.00$\times$) \\
  \textbf{Matrix Mult.}
    & 13.21 & 25.40 (1.92$\times$) & 25.72 (1.95$\times$) \\
  \bottomrule
  \end{tabular}%
  }
  \caption{Bounds-check hoisting pass runtimes in seconds on AMD. Parentheses compare against C as the baseline.}
  \label{tab:sli-runtime}
\end{minipage}

\end{table}

\section{Discussion}

\sysname demonstrates that the performance of memory-safe systems languages is a dynamic interplay between 
the costs of the safety checks, the
compiler-level optimizations,
and microarchitectural speculation. 
Future language designs and transpilation systems can leverage this 
interplay to reconcile safety and performance without manual tuning.
The cases where the performance of Rust code falls short reveal opportunities
for further work. 

\para{Performance Modeling Challenges.}
Our results show that C-to-Rust transpilation overhead cannot be modeled as a uniform percentage increase. 
The Rust overhead varies substantially with workload characteristics such as data access and branch behavior. 
Traditional models that assume stable relationships between abstraction boundaries and microarchitectural events~\cite{van2016analytical,cabezas2014extending}
are therefore insufficient, and cannot
accurately predict the performance cost.
This necessitates the development of
novel performance models that can accurately
predict the performance of the transpiled 
Rust code from the C version.

\para{Hardware Sensitivity and Generalizability.} 
Our evaluation includes two widely deployed server-class microarchitectures—AMD EPYC and Intel Xeon Silver—which together represent the majority of modern datacenter and HPC deployments~\cite{mercury2025server, top500nov2024}. 
Across both platforms, the key trends hold. Rust implementations that match or outperform C on one architecture generally do so on the other, and the microarchitectural mechanisms responsible 
are common to both vendors' designs. This consistency suggests that our findings are likely to generalize to other modern processors.
Nonetheless, the magnitude of overhead varies between the two machines, confirming that microarchitectural features such as cache organization, branch predictor design, and out-of-order execution width influence the relative cost of runtime safety checks~\cite{skadron2002branch, nagarakatte2013hardware}. Consequently, the performance ordering between competing implementation strategies may shift across hardware, undermining the notion of a single "best" translation. 

\para{Implications for Transpiler Toolchains.}
The lack of a definitive superior implementation strategy complicates
the development of automated C-to-Rust transpiler toolchains. 
Present-day C-to-Rust transpilation 
toolchains~\cite{hanliangownership,emre2023aliasing,emre_translating_2021} 
perform a line-by-line
transpilation from C to Rust. Such systems would struggle to transpile
C-style imperative code to idiomatic
Rust paradigms. 
More recently, LLM-based transpiler toolchains~\cite{yang_vert_2024, eniser_towards_2024, rustassure}
have proved modestly
successful in transpiling C code to \emph{idiomatic} Rust, but they struggle
with correctness~\cite{li2024translating, yang_vert_2024,farrukh2025safetrans}. 
Eliciting performant output from these
approaches compounds the challenge.

\section{Related Works}
\label{sec:related-works}

Several studies have attempted to examined the performance gap between C and Rust.
Popescu et al.~\cite{popescu2021safer} showed that developers struggle to identify costly bounds checks, and proposed NADER to selectively reintroduce them under a performance budget.
Zhang et al.~\cite{zhang2022towards} found that Rust benchmarks, when implemented in a C-like
manner, generally incur higher overhead than C, due to run-time checks and language constraints.
Fromherz and Protzenko~\cite{fromherz2024compiling} observed that Rust can outperform unoptimized C but lags behind optimized C and that inlining causes significant gains, while removing bounds checks has little effect. Licona-Luque et al.~\cite{rust-wo-shortcuts} found that hand-written safe Rust achieves performance comparable to C across a range of benchmarks, but that relative instruction counts can vary substantially across hardware platforms.
Li et al.~\cite{li2024empirical} reported inconsistent overheads for Rust in Linux drivers, citing coarse-grained locking and array bounds checks as key factors causing degradation, and also noted higher cache/TLB/branch miss rates due to Rust’s ownership patterns.
Together, these studies highlight the need for a deeper systematic performance analysis
of Rust code translated from C. 

\para{AI-assisted Debugging.}
Recent work has explored using large language models (LLMs) for performance analysis and optimization.
POLO~\cite{bai2025polo} and SysLLMatic~\cite{peng2026sysllmatic} use profiling-guided LLM agents to identify performance hotspots and optimize code.
Other work focuses on performance diagnosis. Gigiprofiler~\cite{hu2026gigiprofiler} combines LLM reasoning with static analysis and runtime tracking to identify bottlenecks, while Argus~\cite{nitu2026argus} and AgentDebug~\cite{panahandeh2026rethinking} use agentic reasoning and runtime evidence to localize performance issues.
Our work differs by explaining performance differences between equivalent C and Rust implementations.
We develop a multi-stage agent pipeline that combines performance counters, compiler remarks, and disassembly to investigate differences at three levels: programming idioms, compiler optimizations, and microarchitectural behavior.
This evidence-driven approach attributes performance differences to their underlying causes rather than simply identifying bottlenecks or generating optimizations.

\para{Compiler Optimization for Runtime Memory Safety Checks}
Moving bounds checks out of loops is not a new idea. Gupta~\cite{gupta1990fresh,gupta1993optimizing} and Kolte and
Wolfe~\cite{kolte1995elimination} hoist checks on induction-variable indices
into the preheader, which reports an error before the loop instead of at the
faulting access. Java's precise exceptions forbid that, so Midkiff et
al.~\cite{midkiff1998optimizing} split a loop's iterations into regions that
need checks and regions that do not, versioning the loop on a runtime test.
Both work for indices that are affine in the loop index.
ABCD~\cite{bodik2000abcd} and later work~\cite{yang2009array} remove only checks
they prove redundant at compile time, with no runtime test. JIT compilers hoist
or widen checks and recover through deoptimization when a hoisted check fails,
as in HotSpot~\cite{wurthinger2007array} and LLVM's
LoopPredication~\cite{llvm-looppred}. Rust is compiled ahead of time and has no
deoptimization, so our pass keeps the original loop as the fallback, as Midkiff
et al.\ do. What is new is the set of checks it can prove. Rules~2--4 handle
indices that are not affine functions of the loop counter: indices that are conditionally selected, counters that advance only on some paths, and indices bounded by a guard condition.

LLVM's inductive range check elimination (IRCE)~\cite{llvm-irce} is the closest existing pass. Like
Midkiff et al., it splits a loop into pre-, main and post-loops and removes checks only from the
main loop. It computes the main loop's range by intersecting each check's safe range with the
loop's own iteration range, which imposes three requirements: 1) the index must be an affine function
of an induction variable with a constant step 2) the bound must be loop invariant and 3) the loop must
exit on that same induction variable. None of the hot checks in our benchmarks meets all three of these requirements, and it therefore could not
improve the performance of any of our benchmarks.

\section{Conclusion}

We showed that Rust's safety checks
are not always the dominant source of overhead, as compiler
and microarchitectural optimizations often mitigate their costs.
Building on these findings, we presented \sysname, an AI-assisted
performance debugger that identifies cross-layer performance
bottlenecks and guides safety-preserving compiler optimizations.
Together, these contributions provide a systematic approach
to understanding, diagnosing, and reducing performance
overheads in C-to-Rust migration.

\section{AI Use}

\sysname is itself an AI-assisted tool. An LLM coding agent carries out each of its
three phases (\autoref{sec:ferroscope}), and the verdicts and diagnoses in
\autoref{sec:eval} are its outputs. We checked each diagnosis
against our manual analysis of the same benchmark.
We used Anthropic's Claude Code to write the script that plots
\autoref{fig:runtime_overhead}, and we verified the values against the tables programmatically. The authors take full responsibility for the contents of this paper.

\bibliographystyle{IEEEtranS}
\bibliography{main}
\newpage
\appendix

\section{Intel vs. AMD Instruction-Mix Differences}
\label{app:inst-mix-diff}
As shown in \autoref{tab:ratios_amd} and 
\autoref{tab:ratios_intel}, 
we observe notable differences in dynamic instruction-mix
between the Intel and AMD architectures, particularly in store instructions.
We used a Pin tool to further analyze the instruction mix and found that in BFS, the Intel platform exhibits a 47\% increase in store instructions over AMD for the C implementation and a 60\% increase for both Rust and \rmm{}. Merge Sort (TDR) shows a similar pattern, with increases of 16\%, 38\%, and 64\% for C, Rust, and \rmm{} respectively. PageRank presents a more nuanced case: Intel shows an approximately 40\% increase in stores for both C and \rmm{} but produces a nearly identical count for standard Rust.
We attribute these differences primarily to 
architecture-specific optimizations within \emph{glibc}, which uses runtime dispatching, via \texttt{ifunc}, to select architecture-tuned implementations of 
common functions such as \texttt{memset} and \texttt{memcpy}. 
For instance, the \texttt{glibc} implementation 
on Intel may favor repeated \texttt{STOSB} instructions, leveraging the Enhanced REP MOVSB/STOSB capability present on recent Intel processors, whereas the AMD-targeted variant may instead employ wider vectorized \texttt{MOV} instructions. Because a single vectorized store writes more data per instruction than a byte-granularity \texttt{STOSB}, the same  operation can produce significantly 
different dynamic instruction count without altering the performance.

\end{document}